
\magnification=\magstephalf

\newbox\SlashedBox 
\def\slashed#1{\setbox\SlashedBox=\hbox{#1}
\hbox to 0pt{\hbox to 1\wd\SlashedBox{\hfil/\hfil}\hss}{#1}}
\def\hboxtosizeof#1#2{\setbox\SlashedBox=\hbox{#1}
\hbox to 1\wd\SlashedBox{#2}}

\def\mathslashed#1{\setbox\SlashedBox=\hbox{$#1$}
\hbox to 0pt{\hbox to 1\wd\SlashedBox{\hfil/\hfil}\hss}#1}

\def\ifsmall{\iffalse}  
\def\titlepagefont{}  

\def\DefineTeXgraphics{%
\special{ps::[global] /TeXgraphics { } def}}  

\def\today{\ifcase\month\or January\or February\or March\or April\or May
\or June\or July\or August\or September\or October\or November\or
December\fi\space\number\day, \number\year}
\def\eatPrefix19{}
\def\Year{\expandafter\eatPrefix\the\year}
\newcount\hours \newcount\minutes
\def\monthname{\ifcase\month\or
January\or February\or March\or April\or May\or June\or July\or
August\or September\or October\or November\or December\fi}
\def\shortmonthname{\ifcase\month\or
Jan\or Feb\or Mar\or Apr\or May\or Jun\or Jul\or
Aug\or Sep\or Oct\or Nov\or Dec\fi}

\def\TimeStamp{\hours\the\time\divide\hours by60%
\minutes -\the\time\divide\minutes by60\multiply\minutes by60%
\advance\minutes by\the\time%
${\rm \shortmonthname}\cdot\if\day<10{}0\fi\the\day\cdot\the\year%
\qquad\the\hours:\if\minutes<10{}0\fi\the\minutes$}




\def\Title#1{%
\vskip 1in{\titlefont\centerline{#1}}\vskip .5in}
 
\def\Date#1{\leftline{#1}\tenrm\supereject%
\global\hsize=\hsbody\global\hoffset=\hbodyoffset%
\footline={\hss\tenrm\folio\hss}}

\newif\ifdraftmode
\newif\ifleftlabels  

\def\nolabels{\def\wrlabeL##1{}\def\eqlabeL##1{}\def\reflabeL##1{}}
\def\writelabels{\def\wrlabeL##1{\leavevmode\vadjust{\rlap{\smash%
{\line{{\escapechar=` \hfill\rlap{\sevenrm\hskip.03in\string##1}}}}}}}%
\def\eqlabeL##1{{\escapechar-1\rlap{\sevenrm\hskip.05in\string##1}}}%
\def\reflabeL##1{\noexpand\rlap{\noexpand\sevenrm[\string##1]}}}
\def\writeleftlabels{\def\wrlabeL##1{\leavevmode\vadjust{\rlap{\smash%
{\line{{\escapechar=` \hfill\rlap{\sevenrm\hskip.03in\string##1}}}}}}}%
\def\eqlabeL##1{{\escapechar-1%
\rlap{\sixrm\hskip.05in\string##1}%
\llap{\sevenrm\string##1\hskip.03in\hbox to \hsize{}}}}%
\def\reflabeL##1{\noexpand\rlap{\noexpand\sevenrm[\string##1]}}}
\nolabels



\newdimen\fullhsize
\newdimen\hstitle
\hstitle=\hsize 
\newdimen\hsbody
\hsbody=\hsize 
\newdimen\hbodyoffset
\hbodyoffset=\hoffset 
\newbox\leftpage
\def\abstract#1{#1}
\def\rotated{\special{ps: landscape}
\magnification=1000  
\baselineskip=14pt
\global\hstitle=9truein\global\hsbody=4.75truein
\global\vsize=7truein\global\voffset=-.31truein
\global\hoffset=-0.54in\global\hbodyoffset=-.54truein
\global\fullhsize=10truein
\def\DefineTeXgraphics{%
\special{ps::[global] 
/TeXgraphics {currentpoint translate 0.7 0.7 scale
              -80 0.72 mul -1000 0.72 mul translate} def}}
\let\lr=L
\def\ifsmall{\iftrue}
\def\titlepagefont{\twelvepoint}
\trueseventeenpoint
\def\almostshipout##1{\if L\lr \count1=1
      \global\setbox\leftpage=##1 \global\let\lr=R
   \else \count1=2
      \shipout\vbox{\hbox to\fullhsize{\box\leftpage\hfil##1}}
      \global\let\lr=L\fi}

\output={\ifnum\count0=1 
 \shipout\vbox{\hbox to \fullhsize{\hfill\pagebody\hfill}}\advancepageno
 \else
 \almostshipout{\leftline{\vbox{\pagebody\makefootline}}}\advancepageno 
 \fi}

\def\abstract##1{{\leftskip=1.5in\rightskip=1.5in ##1\par}} }

\def\linemessage#1{\immediate\write16{#1}}

\global\newcount\secno \global\secno=0
\global\newcount\appno \global\appno=0
\global\newcount\meqno \global\meqno=1
\global\newcount\subsecno \global\subsecno=0
\global\newcount\figno \global\figno=0

\newif\ifAnyCounterChanged
\let\terminator=\relax
\def\normalize#1{\ifx#1\terminator\let\next=\relax\else%
\if#1i\aftergroup i\else\if#1v\aftergroup v\else\if#1x\aftergroup x%
\else\if#1l\aftergroup l\else\if#1c\aftergroup c\else%
\if#1m\aftergroup m\else%
\if#1I\aftergroup I\else\if#1V\aftergroup V\else\if#1X\aftergroup X%
\else\if#1L\aftergroup L\else\if#1C\aftergroup C\else%
\if#1M\aftergroup M\else\aftergroup#1\fi\fi\fi\fi\fi\fi\fi\fi\fi\fi\fi\fi%
\let\next=\normalize\fi%
\next}
\def\makeNormal#1#2{\def\doNormalDef{\edef#1}\begingroup%
\aftergroup\doNormalDef\aftergroup{\normalize#2\terminator\aftergroup}%
\endgroup}

\def\warnIfChanged#1#2{%
\ifundef#1
\else\begingroup%
\edef\oldDefinitionOfCounter{#1}\edef\newDefinitionOfCounter{#2}%
\ifx\oldDefinitionOfCounter\newDefinitionOfCounter%
\else%
\linemessage{Warning: definition of \noexpand#1 has changed.}%
\global\AnyCounterChangedtrue\fi\endgroup\fi}

\def\Section#1{\global\advance\secno by1\relax\global\meqno=1%
\global\subsecno=0%
\bigbreak\bigskip
\centerline{\twelvepoint \bf %
\the\secno. #1}%
\par\nobreak\medskip\nobreak}
\def\tagsection#1{%
\warnIfChanged#1{\the\secno}%
\xdef#1{\the\secno}%
\ifWritingAuxFile\immediate\write\auxfile{\noexpand\xdef\noexpand#1{#1}}\fi%
}
\def\section{\Section}
\def\Subsection#1{\global\advance\subsecno by1\relax\medskip %
\leftline{\bf\the\secno.\the\subsecno\ #1}%
\par\nobreak\smallskip\nobreak}
\def\tagsubsection#1{%
\warnIfChanged#1{\the\secno.\the\subsecno}%
\xdef#1{\the\secno.\the\subsecno}%
\ifWritingAuxFile\immediate\write\auxfile{\noexpand\xdef\noexpand#1{#1}}\fi%
}

\def\subsection{\Subsection}

\def\romappno{\uppercase\expandafter{\romannumeral\appno}}
\def\makeNormalizedRomappno{%
\expandafter\makeNormal\expandafter\normalizedromappno%
\expandafter{\romannumeral\appno}%
\edef\normalizedromappno{\uppercase{\normalizedromappno}}}
\def\Appendix#1{\global\advance\appno by1\relax\global\meqno=1\global\secno=0%
\global\subsecno=0%
\bigbreak\bigskip
\centerline{\twelvepoint \bf Appendix %
\romappno. #1}%
\par\nobreak\medskip\nobreak}
\def\tagappendix#1{\makeNormalizedRomappno%
\warnIfChanged#1{\normalizedromappno}%
\xdef#1{\normalizedromappno}%
\ifWritingAuxFile\immediate\write\auxfile{\noexpand\xdef\noexpand#1{#1}}\fi%
}
\def\appendix{\Appendix}
\def\Subappendix#1{\global\advance\subsecno by1\relax\medskip %
\leftline{\bf\romappno.\the\subsecno\ #1}%
\par\nobreak\smallskip\nobreak}
\def\tagsubappendix#1{\makeNormalizedRomappno%
\warnIfChanged#1{\normalizedromappno.\the\subsecno}%
\xdef#1{\normalizedromappno.\the\subsecno}%
\ifWritingAuxFile\immediate\write\auxfile{\noexpand\xdef\noexpand#1{#1}}\fi%
}

\def\eqn#1{\makeNormalizedRomappno%
\ifnum\secno>0%
  \warnIfChanged#1{\the\secno.\the\meqno}%
  \eqno(\the\secno.\the\meqno)\xdef#1{\the\secno.\the\meqno}%
     \global\advance\meqno by1
\else\ifnum\appno>0%
  \warnIfChanged#1{\normalizedromappno.\the\meqno}%
  \eqno({\rm\romappno}.\the\meqno)%
      \xdef#1{\normalizedromappno.\the\meqno}%
     \global\advance\meqno by1
\else%
  \warnIfChanged#1{\the\meqno}%
  \eqno(\the\meqno)\xdef#1{\the\meqno}%
     \global\advance\meqno by1
\fi\fi%
\eqlabeL#1%
\ifWritingAuxFile\immediate\write\auxfile{\noexpand\xdef\noexpand#1{#1}}\fi%
}
\def\defeqn#1{\makeNormalizedRomappno%
\ifnum\secno>0%
  \warnIfChanged#1{\the\secno.\the\meqno}%
  \xdef#1{\the\secno.\the\meqno}%
     \global\advance\meqno by1
\else\ifnum\appno>0%
  \warnIfChanged#1{\normalizedromappno.\the\meqno}%
  \xdef#1{\normalizedromappno.\the\meqno}%
     \global\advance\meqno by1
\else%
  \warnIfChanged#1{\the\meqno}%
  \xdef#1{\the\meqno}%
     \global\advance\meqno by1
\fi\fi%
\eqlabeL#1%
\ifWritingAuxFile\immediate\write\auxfile{\noexpand\xdef\noexpand#1{#1}}\fi%
}
\def\anoneqn{\makeNormalizedRomappno%
\ifnum\secno>0
  \eqno(\the\secno.\the\meqno)%
     \global\advance\meqno by1
\else\ifnum\appno>0
  \eqno({\rm\normalizedromappno}.\the\meqno)%
     \global\advance\meqno by1
\else
  \eqno(\the\meqno)%
     \global\advance\meqno by1
\fi\fi%
}
\def\mfig#1#2{\ifx#20
\else\global\advance\figno by1%
\relax#1\the\figno%
\warnIfChanged#2{\the\figno}%
\xdef#2{\the\figno}%
\reflabeL#2%
\ifWritingAuxFile\immediate\write\auxfile{\noexpand\xdef\noexpand#2{#2}}\fi\fi%
}

\def\fig#1{\mfig{fig.\ ~}#1}

\catcode`@=11 

\newif\ifFiguresInText\FiguresInTexttrue
\newif\if@FigureFileCreated
\newwrite\capfile
\newwrite\figfile

\newif\ifcaption
\captiontrue
\def\captionsize{\tenrm}
\def\PlaceTextFigure#1#2#3#4{%
\vskip 0.5truein%
#3\hfil\epsfbox{#4}\hfil\break%
\ifcaption\hfil\vbox{\captionsize Figure #1. #2}\hfil\fi%
\vskip10pt}
\def\PlaceEndFigure#1#2{%
\epsfxsize=\hsize\epsfbox{#2}\vfill\centerline{Figure #1.}\eject}

\def\LoadFigure#1#2#3#4{%
\ifundef#1{\phantom{\mfig{}#1}}\else
\fi%
\ifFiguresInText
\PlaceTextFigure{#1}{#2}{#3}{#4}%
\else
\if@FigureFileCreated\else%
\immediate\openout\capfile=\jobname.caps%
\immediate\openout\figfile=\jobname.figs%
@FigureFileCreatedtrue\fi%
\immediate\write\capfile{\noexpand\item{Figure \noexpand#1.\ }{#2}\vskip10pt}%
\immediate\write\figfile{\noexpand\PlaceEndFigure\noexpand#1{\noexpand#4}}%
\fi}

\def\listfigs{\ifFiguresInText\else%
\vfill\eject\immediate\closeout\capfile
\immediate\closeout\figfile%
\centerline{{\bf Figures}}\bigskip\frenchspacing%
\catcode`@=11 
\def\captionsize{\tenrm}
\input \jobname.caps\vfill\eject\nonfrenchspacing%
\catcode`\@=\active
\catcode`@=12  
\input\jobname.figs\fi}

\font\ninerm=cmr9
\font\eightrm=cmr8
\font\sixrm=cmr6

\def\loadtrueseventeenpoint{
 \font\seventeenrm=cmr10 at 17.28truept
 \font\seventeeni=cmmi10 at 17.28truept
 \font\seventeenbf=cmbx10 at 17.28truept
 \font\seventeenit=cmti10 at 17.28truept
 \font\seventeensl=cmsl10 at 17.28truept
 \font\seventeensy=cmsy10 at 17.28truept
}
\def\loadfourteenpoint{
\font\fourteenrm=cmr10 at 14.4pt
\font\fourteeni=cmmi10 at 14.4pt
\font\fourteenit=cmti10 at 14.4pt
\font\fourteensl=cmsl10 at 14.4pt
\font\fourteensy=cmsy10 at 14.4pt
\font\fourteenbf=cmbx10 at 14.4pt
}
\def\loadtruetwelvepoint{
\font\twelverm=cmr10 at 12truept
\font\twelvei=cmmi10 at 12truept
\font\twelveit=cmti10 at 12truept
\font\twelvesl=cmsl10 at 12truept
\font\twelvesy=cmsy10 at 12truept
\font\twelvebf=cmbx10 at 12truept
}

\font\ninei=cmmi9
\font\eighti=cmmi8
\font\sixi=cmmi6
\skewchar\ninei='177 \skewchar\eighti='177 \skewchar\sixi='177

\font\ninesy=cmsy9
\font\eightsy=cmsy8
\font\sixsy=cmsy6
\skewchar\ninesy='60 \skewchar\eightsy='60 \skewchar\sixsy='60

\font\ninebf=cmbx9
\font\eightbf=cmbx8
\font\sixbf=cmbx6

\font\ninett=cmtt9
\font\eighttt=cmtt8

\hyphenchar\tentt=-1 
\hyphenchar\ninett=-1
\hyphenchar\eighttt=-1         

\font\ninesl=cmsl9
\font\eightsl=cmsl8

\font\nineit=cmti9
\font\eightit=cmti8

                      
\newskip\ttglue
\def\tenpoint{\def\rm{\fam0\tenrm}%
  \textfont0=\tenrm \scriptfont0=\sevenrm \scriptscriptfont0=\fiverm
  \textfont1=\teni \scriptfont1=\seveni \scriptscriptfont1=\fivei
  \textfont2=\tensy \scriptfont2=\sevensy \scriptscriptfont2=\fivesy
  \textfont3=\tenex \scriptfont3=\tenex \scriptscriptfont3=\tenex
  \def\it{\fam\itfam\tenit}\textfont\itfam=\tenit
  \def\sl{\fam\slfam\tensl}\textfont\slfam=\tensl
  \def\bf{\fam\bffam\tenbf}\textfont\bffam=\tenbf \scriptfont\bffam=\sevenbf
  \scriptscriptfont\bffam=\fivebf
  \normalbaselineskip=12pt
  \let\sc=\eightrm
  \let\big=\tenbig
  \setbox\strutbox=\hbox{\vrule height8.5pt depth3.5pt width\z@}%
  \normalbaselines\rm}

\def\twelvepoint{\def\rm{\fam0\twelverm}%
  \textfont0=\twelverm \scriptfont0=\ninerm \scriptscriptfont0=\sevenrm
  \textfont1=\twelvei \scriptfont1=\ninei \scriptscriptfont1=\seveni
  \textfont2=\twelvesy \scriptfont2=\ninesy \scriptscriptfont2=\sevensy
  \textfont3=\tenex \scriptfont3=\tenex \scriptscriptfont3=\tenex
  \def\it{\fam\itfam\twelveit}\textfont\itfam=\twelveit
  \def\sl{\fam\slfam\twelvesl}\textfont\slfam=\twelvesl
  \def\bf{\fam\bffam\twelvebf}\textfont\bffam=\twelvebf%
  \scriptfont\bffam=\ninebf
  \scriptscriptfont\bffam=\sevenbf
  \normalbaselineskip=12pt
  \let\sc=\eightrm
  \let\big=\tenbig
  \setbox\strutbox=\hbox{\vrule height8.5pt depth3.5pt width\z@}%
  \normalbaselines\rm}

\def\fourteenpoint{\def\rm{\fam0\fourteenrm}%
  \textfont0=\fourteenrm \scriptfont0=\tenrm \scriptscriptfont0=\sevenrm
  \textfont1=\fourteeni \scriptfont1=\teni \scriptscriptfont1=\seveni
  \textfont2=\fourteensy \scriptfont2=\tensy \scriptscriptfont2=\sevensy
  \textfont3=\tenex \scriptfont3=\tenex \scriptscriptfont3=\tenex
  \def\it{\fam\itfam\fourteenit}\textfont\itfam=\fourteenit
  \def\sl{\fam\slfam\fourteensl}\textfont\slfam=\fourteensl
  \def\bf{\fam\bffam\fourteenbf}\textfont\bffam=\fourteenbf%
  \scriptfont\bffam=\tenbf
  \scriptscriptfont\bffam=\sevenbf
  \normalbaselineskip=17pt
  \let\sc=\elevenrm
  \let\big=\tenbig                                          
  \setbox\strutbox=\hbox{\vrule height8.5pt depth3.5pt width\z@}%
  \normalbaselines\rm}

\def\seventeenpoint{\def\rm{\fam0\seventeenrm}%
  \textfont0=\seventeenrm \scriptfont0=\fourteenrm \scriptscriptfont0=\tenrm
  \textfont1=\seventeeni \scriptfont1=\fourteeni \scriptscriptfont1=\teni
  \textfont2=\seventeensy \scriptfont2=\fourteensy \scriptscriptfont2=\tensy
  \textfont3=\tenex \scriptfont3=\tenex \scriptscriptfont3=\tenex
  \def\it{\fam\itfam\seventeenit}\textfont\itfam=\seventeenit
  \def\sl{\fam\slfam\seventeensl}\textfont\slfam=\seventeensl
  \def\bf{\fam\bffam\seventeenbf}\textfont\bffam=\seventeenbf%
  \scriptfont\bffam=\fourteenbf
  \scriptscriptfont\bffam=\twelvebf
  \normalbaselineskip=21pt
  \let\sc=\fourteenrm
  \let\big=\tenbig                                          
  \setbox\strutbox=\hbox{\vrule height 12pt depth 6pt width\z@}%
  \normalbaselines\rm}

\def\ninepoint{\def\rm{\fam0\ninerm}%
  \textfont0=\ninerm \scriptfont0=\sixrm \scriptscriptfont0=\fiverm
  \textfont1=\ninei \scriptfont1=\sixi \scriptscriptfont1=\fivei
  \textfont2=\ninesy \scriptfont2=\sixsy \scriptscriptfont2=\fivesy
  \textfont3=\tenex \scriptfont3=\tenex \scriptscriptfont3=\tenex
  \def\it{\fam\itfam\nineit}\textfont\itfam=\nineit
  \def\sl{\fam\slfam\ninesl}\textfont\slfam=\ninesl
  \def\bf{\fam\bffam\ninebf}\textfont\bffam=\ninebf \scriptfont\bffam=\sixbf
  \scriptscriptfont\bffam=\fivebf
  \normalbaselineskip=11pt
  \let\sc=\sevenrm
  \let\big=\ninebig
  \setbox\strutbox=\hbox{\vrule height8pt depth3pt width\z@}%
  \normalbaselines\rm}

\def\eightpoint{\def\rm{\fam0\eightrm}%
  \textfont0=\eightrm \scriptfont0=\sixrm \scriptscriptfont0=\fiverm%
  \textfont1=\eighti \scriptfont1=\sixi \scriptscriptfont1=\fivei%
  \textfont2=\eightsy \scriptfont2=\sixsy \scriptscriptfont2=\fivesy%
  \textfont3=\tenex \scriptfont3=\tenex \scriptscriptfont3=\tenex%
  \def\it{\fam\itfam\eightit}\textfont\itfam=\eightit%
  \def\sl{\fam\slfam\eightsl}\textfont\slfam=\eightsl%
  \def\bf{\fam\bffam\eightbf}\textfont\bffam=\eightbf \scriptfont\bffam=\sixbf%
  \scriptscriptfont\bffam=\fivebf%
  \normalbaselineskip=9pt%
  \let\sc=\sixrm%
  \let\big=\eightbig%
  \setbox\strutbox=\hbox{\vrule height7pt depth2pt width\z@}%
  \normalbaselines\rm}

\def\tenbig#1{{\hbox{$\left#1\vbox to8.5pt{}\right.\n@space$}}}
\def\ninebig#1{{\hbox{$\textfont0=\tenrm\textfont2=\tensy
  \left#1\vbox to7.25pt{}\right.\n@space$}}}
\def\eightbig#1{{\hbox{$\textfont0=\ninerm\textfont2=\ninesy
  \left#1\vbox to6.5pt{}\right.\n@space$}}}

\def\footnote#1{\edef\@sf{\spacefactor\the\spacefactor}#1\@sf
      \insert\footins\bgroup\eightpoint
      \interlinepenalty100 \let\par=\endgraf
        \leftskip=\z@skip \rightskip=\z@skip
        \splittopskip=10pt plus 1pt minus 1pt \floatingpenalty=20000
        \smallskip\item{#1}\bgroup\strut\aftergroup\@foot\let\next}
\skip\footins=12pt plus 2pt minus 4pt 
\dimen\footins=30pc 

\newinsert\margin
\dimen\margin=\maxdimen
\def\titlefont{\seventeenpoint}
\loadtruetwelvepoint 
\loadtrueseventeenpoint

\def\eatOne#1{}
\def\ifundef#1{\expandafter\ifx%
\csname\expandafter\eatOne\string#1\endcsname\relax}
\def\notTrue{\iffalse}\def\isTrue{\iftrue}
\def\ifdef#1{{\ifundef#1%
\aftergroup\notTrue\else\aftergroup\isTrue\fi}}
\def\use#1{\ifundef#1\linemessage{Warning: \string#1 is undefined.}%
{\tt \string#1}\else#1\fi}



%
\catcode`"=11
\let\quote="
\catcode`"=12
\chardef\foo="22
\global\newcount\refno \global\refno=1
\newwrite\rfile
\newlinechar=`\^^J
\def\@ref#1#2{\the\refno\n@ref#1{#2}}
\def\h@ref#1#2#3{\href{#3}{\the\refno}\n@ref#1{#2}}
\def\n@ref#1#2{\xdef#1{\the\refno}%
\ifnum\refno=1\immediate\openout\rfile=\jobname.refs\fi%
\immediate\write\rfile{\noexpand\item{[\noexpand#1]\ }#2.}%
\global\advance\refno by1}
\def\nref{\n@ref} 
\def\ref{\@ref}   
\def\hrref{\h@ref}
\def\lref#1#2{\the\refno\xdef#1{\the\refno}%
\ifnum\refno=1\immediate\openout\rfile=\jobname.refs\fi%
\immediate\write\rfile{\noexpand\item{[\noexpand#1]\ }#2\semi}%
\global\advance\refno by1}
\def\cref#1{\immediate\write\rfile{#1\semi}}

\def\preref#1#2{\gdef#1{\@ref#1{#2}}}

\def\semi{;\hfil\noexpand\break}

\def\listrefs{\vfill\eject\immediate\closeout\rfile
\centerline{{\bf References}}\bigskip\frenchspacing%
\input \jobname.refs\vfill\eject\nonfrenchspacing}

\def\inputAuxIfPresent#1{\immediate\openin1=#1
\ifeof1\message{No file \auxfileName; I'll create one.
}\else\closein1\relax\input\auxfileName\fi%
}
\def\NPB{Nucl.\ Phys.\ B}
\def\PRL{Phys.\ Rev.\ Lett.\ }




\newif\ifWritingAuxFile
\newwrite\auxfile
\def\SetUpAuxFile{%
\xdef\auxfileName{\jobname.aux}%
\inputAuxIfPresent{\auxfileName}%
\WritingAuxFiletrue%
\immediate\openout\auxfile=\auxfileName}

\def\L{\left(}\def\R{\right)}


\catcode`\@=\active
\catcode`@=12  
\catcode`\"=\active


\def\L{\left(}
\def\R{\right)}

\def\s{{1\over6}}

\def\Re{\mathop{\rm Re}\nolimits}
\def\Im{\mathop{\rm Im}\nolimits}

\def\eps{\epsilon}

\def\d{d^{\vphantom{\dagger}}}
\def\pol{\varepsilon}

\def\dl^#1_#2{\delta^{#1}{}_{#2}}

\def\Li{\mathop{\rm Li}\nolimits}

\def\Ord{{\cal O}}

\def\A#1{{\cal A}_{#1}}

\catcode`@=11  
\def\meqalign#1{\,\vcenter{\openup1\jot\m@th
   \ialign{\strut\hfil$\displaystyle{##}$ && $\displaystyle{{}##}$\hfil
             \crcr#1\crcr}}\,}
\catcode`@=12  


\baselineskip 15pt
\overfullrule 0.5pt


\def\A#1{{\cal A}_{#1}}

\def\pol{\varepsilon}

\def\ksl{\slashed{k}}

\def\eb{\bar{\eta}}

\def\Re{\mathop{\rm Re}}
\def\L{\left(}\def\R{\right)}

\def\spa#1.#2{\left\langle#1\,#2\right\rangle}
\def\spb#1.#2{\left[#1\,#2\right]}
\def\lor#1.#2{\left(#1\,#2\right)}
\def\sand#1.#2.#3{%
\left\langle\smash{#1}{\vphantom1}^{-}\right|{#2}%
\left|\smash{#3}{\vphantom1}^{-}\right\rangle}
\def\sandp#1.#2.#3{%
\left\langle\smash{#1}{\vphantom1}^{-}\right|{#2}%
\left|\smash{#3}{\vphantom1}^{+}\right\rangle}
\def\sandpp#1.#2.#3{%
\left\langle\smash{#1}{\vphantom1}^{+}\right|{#2}%
\left|\smash{#3}{\vphantom1}^{+}\right\rangle}
\def\sandpm#1.#2.#3{%
\left\langle\smash{#1}{\vphantom1}^{+}\right|{#2}%
\left|\smash{#3}{\vphantom1}^{-}\right\rangle}
\def\sandmp#1.#2.#3{%
\left\langle\smash{#1}{\vphantom1}^{-}\right|{#2}%
\left|\smash{#3}{\vphantom1}^{+}\right\rangle}
\catcode`@=11  
\def\meqalign#1{\,\vcenter{\openup1\jot\m@th
   \ialign{\strut\hfil$\displaystyle{##}$ && $\displaystyle{{}##}$\hfil
             \crcr#1\crcr}}\,}
\catcode`@=12  

\newread\epsffilein    
\newif\ifepsffileok    
\newif\ifepsfbbfound   
\newif\ifepsfverbose   
\newdimen\epsfxsize    
\newdimen\epsfysize    
\newdimen\epsftsize    
\newdimen\epsfrsize    
\newdimen\epsftmp      
\newdimen\pspoints     
\pspoints=1bp          
\epsfxsize=0pt         
\epsfysize=0pt         
\def\epsfbox#1{\global\def\epsfllx{72}\global\def\epsflly{72}%
   \global\def\epsfurx{540}\global\def\epsfury{720}%
   \def\lbracket{[}\def\testit{#1}\ifx\testit\lbracket
   \let\next=\epsfgetlitbb\else\let\next=\epsfnormal\fi\next{#1}}%
\def\epsfgetlitbb#1#2 #3 #4 #5]#6{\epsfgrab #2 #3 #4 #5 .\\%
   \epsfsetgraph{#6}}%
\def\epsfnormal#1{\epsfgetbb{#1}\epsfsetgraph{#1}}%
\def\epsfgetbb#1{%
%
%
\openin\epsffilein=#1
\ifeof\epsffilein\errmessage{I couldn't open #1, will ignore it}\else
%
%
   {\epsffileoktrue \chardef\other=12
    \def\do##1{\catcode`##1=\other}\dospecials \catcode`\ =10
    \loop
       \read\epsffilein to \epsffileline
       \ifeof\epsffilein\epsffileokfalse\else
%
%
          \expandafter\epsfaux\epsffileline:. \\%
       \fi
   \ifepsffileok\repeat
   \ifepsfbbfound\else
    \ifepsfverbose\message{No bounding box comment in #1; using defaults}\fi\fi
   }\closein\epsffilein\fi}%
%
%
\def\epsfclipstring{}
\def\epsfsetgraph#1{%
   \epsfrsize=\epsfury\pspoints
   \advance\epsfrsize by-\epsflly\pspoints
   \epsftsize=\epsfurx\pspoints
   \advance\epsftsize by-\epsfllx\pspoints
%
%
   \epsfxsize\epsfsize\epsftsize\epsfrsize
   \ifnum\epsfxsize=0 \ifnum\epsfysize=0
      \epsfxsize=\epsftsize \epsfysize=\epsfrsize
      \epsfrsize=0pt
%
%
     \else\epsftmp=\epsftsize \divide\epsftmp\epsfrsize
       \epsfxsize=\epsfysize \multiply\epsfxsize\epsftmp
       \multiply\epsftmp\epsfrsize \advance\epsftsize-\epsftmp
       \epsftmp=\epsfysize
       \loop \advance\epsftsize\epsftsize \divide\epsftmp 2
       \ifnum\epsftmp>0
          \ifnum\epsftsize<\epsfrsize\else
             \advance\epsftsize-\epsfrsize \advance\epsfxsize\epsftmp \fi
       \repeat
       \epsfrsize=0pt
     \fi
   \else \ifnum\epsfysize=0
     \epsftmp=\epsfrsize \divide\epsftmp\epsftsize
     \epsfysize=\epsfxsize \multiply\epsfysize\epsftmp   
     \multiply\epsftmp\epsftsize \advance\epsfrsize-\epsftmp
     \epsftmp=\epsfxsize
     \loop \advance\epsfrsize\epsfrsize \divide\epsftmp 2
     \ifnum\epsftmp>0
        \ifnum\epsfrsize<\epsftsize\else
           \advance\epsfrsize-\epsftsize \advance\epsfysize\epsftmp \fi
     \repeat
     \epsfrsize=0pt
    \else
     \epsfrsize=\epsfysize
    \fi
   \fi
%
%
   \ifepsfverbose\message{#1: width=\the\epsfxsize, height=\the\epsfysize}\fi
   \epsftmp=10\epsfxsize \divide\epsftmp\pspoints
   \vbox to\epsfysize{\vfil\hbox to\epsfxsize{%
      \ifnum\epsfrsize=0\relax
        \includegraphics{#1}%
      \else
        \epsfrsize=10\epsfysize \divide\epsfrsize\pspoints
        \includegraphics{#1}%
      \fi
      \hfil}}%
\global\epsfxsize=0pt\global\epsfysize=0pt}%
%
%
{\catcode`\%=12 \global\let\epsfpercent=
%
%
\long\def\epsfaux#1#2:#3\\{\ifx#1\epsfpercent
   \def\testit{#2}\ifx\testit\epsfbblit
      \epsfgrab #3 . . . \\%
      \epsffileokfalse
      \global\epsfbbfoundtrue
   \fi\else\ifx#1\par\else\epsffileokfalse\fi\fi}%
%
%
\def\epsfempty{}%
\def\epsfgrab #1 #2 #3 #4 #5\\{%
\global\def\epsfllx{#1}\ifx\epsfllx\epsfempty
      \epsfgrab #2 #3 #4 #5 .\\\else
   \global\def\epsflly{#2}%
   \global\def\epsfurx{#3}\global\def\epsfury{#4}\fi}%
%
%
\def\epsfsize#1#2{\epsfxsize}
%
%

\SetUpAuxFile
\loadfourteenpoint
\FiguresInTexttrue
\nopagenumbers\hsize=\hstitle\vskip1in
\overfullrule 0pt
\hfuzz 35 pt
\vbadness=10001
%

%

\newbox\charbox
\newbox\slabox
\def\s#1{{      
        \setbox\charbox=\hbox{$#1$}
        \setbox\slabox=\hbox{$/$}
        \dimen\charbox=\ht\slabox
        \advance\dimen\charbox by -\dp\slabox
        \advance\dimen\charbox by -\ht\charbox
        \advance\dimen\charbox by \dp\charbox
        \divide\dimen\charbox by 2
        \raise-\dimen\charbox\hbox to \wd\charbox{\hss/\hss}
        \llap{$#1$}
}}

\def\Li{\mathop{\rm Li}\nolimits}

\def\Ls{\mathop{\rm Ls}\nolimits}
\def\Ll{\mathop{\rm L}\nolimits}

\def\e{\epsilon}


\def\ns{n_{\mskip-2mu s}}
\def\nf{n_{\mskip-2mu f}}
\def\tree{{\rm tree}}
\def\ib{{\; \bar\imath}}

\def\cg{c_\Gamma}

\def\tree{{\rm tree}}

\def\ns{n_{\mskip-2mu s}}\def\nf{n_{\mskip-2mu f}}
\def\hf{{\textstyle{1\over2}}}
\def\lr{\leftrightarrow}
\def\MSbar{$\overline{\rm MS}$}
\def\DRbar{$\overline{\rm DR}$}

\def\spa#1.#2{\left\langle#1\,#2\right\rangle}
\def\spb#1.#2{\left[#1\,#2\right]}
\def\lor#1.#2{\left(#1\,#2\right)}
\def\sand#1.#2.#3{%
  \left\langle\smash{#1}{\vphantom1}\right|{#2}%
  \left|\smash{#3}{\vphantom1}\right\rangle}
\def\sandp#1.#2.#3{%
  \left\langle\smash{#1}{\vphantom1}^{-}\right|{#2}%
  \left|\smash{#3}{\vphantom1}^{+}\right\rangle}
\def\sandpp#1.#2.#3{%
  \left\langle\smash{#1}{\vphantom1}^{+}\right|{#2}%
  \left|\smash{#3}{\vphantom1}^{+}\right\rangle}
\def\sandmm#1.#2.#3{%
  \left\langle\smash{#1}{\vphantom1}^{-}\right|{#2}%
  \left|\smash{#3}{\vphantom1}^{-}\right\rangle}
\def\sandpm#1.#2.#3{%
  \left\langle\smash{#1}{\vphantom1}^{+}\right|{#2}%
  \left|\smash{#3}{\vphantom1}^{-}\right\rangle}
\def\sandmp#1.#2.#3{%
  \left\langle\smash{#1}{\vphantom1}^{-}\right|{#2}%
  \left|\smash{#3}{\vphantom1}^{+}\right\rangle}

\def\spaa#1.#2.#3{\langle\mskip-1mu{#1}^- 
                  | #2 | {#3}^+\mskip-1mu\rangle}
\def\spbb#1.#2.#3{\langle\mskip-1mu{#1}^+ 
                  | #2 | {#3}^-\mskip-1mu\rangle}
\def\spab#1.#2.#3{\langle\mskip-1mu{#1} 
                  | #2 | {#3}\mskip-1mu\rangle}
\def\spba#1.#2.#3{\langle\mskip-1mu{#1}^+ 
                  | #2 | {#3}^+\mskip-1mu\rangle}

\def\"#1{{\accent127 #1}}
\def\ttilde#1{{\accent126 #1}}
%

\def\eb{{\bar e}}
\def\qb{{\bar q}}
\def\Qb{{\bar Q}}
\def\Lsnew{\mathop{\rm \widetilde {Ls}}\nolimits}
\def\d#1#2{\delta_{#1 #2}}

\def\dt{\Delta_3}
\def\rtdelta{\sqrt{\Delta_3}}
\def\rtmdelta{\sqrt{-\Delta_3}}
\def\Cl{{\rm Cl}}
\def\nn{{++}}
\def\an{{+-}}
\def\bn{{+\pm}}
\def\sl{{\rm sl}}
\def\ax{{\rm ax}}
\def\prop#1{{\cal P}_{#1}}

\def\q{{\vphantom{\qb}{q}}}
\def\Q{{\vphantom{\Qb}{Q}}}


\noindent
hep-ph/9610370 \hfill {SLAC--PUB--7316}\break
\rightline{Saclay/SPhT--T96/111}
\rightline{UCLA/96/TEP/33}
\rightline{October, 1996}

\leftlabelstrue
\vskip -.5 in
\Title{One-Loop Amplitudes for $e^+ e^- \rightarrow \qb \q \Qb \Q $}

\centerline{Zvi Bern${}^{\sharp}$}
\baselineskip12truept
\centerline{\it Department of Physics and Astronomy}
\centerline{\it University of California, Los Angeles}
\centerline{\it Los Angeles, CA 90095}
\centerline{\tt bern@physics.ucla.edu}

\smallskip\smallskip

\baselineskip17truept
\centerline{Lance Dixon${}^{\star}$}
\baselineskip12truept
\centerline{\it Stanford Linear Accelerator Center}
\centerline{\it Stanford University}
\centerline{\it Stanford, CA 94309}
\centerline{\tt lance@slac.stanford.edu}


\smallskip\smallskip

\baselineskip17truept
\centerline{David A. Kosower and Stefan Weinzierl}
\baselineskip12truept
\centerline{\it Service de Physique Th\'eorique${}^{\dagger}$}
\centerline{\it Centre d'Etudes de Saclay}
\centerline{\it F-91191 Gif-sur-Yvette cedex, France}
\centerline{\tt kosower@spht.saclay.cea.fr}
\centerline{\tt stefanw@spht.saclay.cea.fr}

\vskip 0.2in\baselineskip13truept

\vskip 0.3truein
\centerline{\bf Abstract}

{\narrower 

We present the one-loop helicity amplitudes for processes involving a
vector boson $V$ ($V=W,Z$, or $\gamma^*$) and four massless quarks, $0
\to V\qb \q\Qb \Q$, where $V$ couples to a massless lepton pair.
These amplitudes are required for next-to-leading order
$\Ord(\alpha_s^3)$ numerical programs for four-jet production at $e^+
e^-$ colliders, for $W$, $Z$ or Drell-Yan production in association
with two jets at hadron colliders, and for three-jet production in
deeply inelastic scattering experiments.  We obtained the amplitudes
presented here by using their analytic properties to constrain their
form. }

\vskip 0.3truein

\centerline{\it Submitted to Nuclear Physics B}

\vfill
\vskip 0.1in
\noindent\hrule width 3.6in\hfil\break
\noindent
${}^{\sharp}$Research supported in part by the US Department of Energy
under grant DE-FG03-91ER40662 and in part by the
Alfred P. Sloan Foundation under grant BR-3222. \hfil\break
${}^{\star}$Research supported by the US Department of
Energy under grant DE-AC03-76SF00515.\hfil\break
${}^{\dagger}$Laboratory of the
{\it Direction des Sciences de la Mati\`ere\/}
of the {\it Commissariat \`a l'Energie Atomique\/} of France.\hfil\break

\Date{}

\line{}

\baselineskip17pt
%


\preref\ThreeJetsBorn{%
J. Ellis, M.K. Gaillard and G.G. Ross, Nucl.\ Phys.\ B111:253 (1976)}

\preref\FourJetsBorn{A. Ali, et al., Phys.\ Lett.\ 82B:285 (1979); 
Nucl.\ Phys.\ B167:454 (1980)}

\preref\FiveJetsBorn{%
K. Hagiwara and D. Zeppenfeld, Nucl. Phys. B313:560 (1989)\semi 
F.A. Berends, W.T. Giele and H. Kuijf, Nucl.\ Phys.\ B321:39
(1989)\semi
N.K. Falk, D. Graudenz and G. Kramer, Nucl. Phys. B328:317 (1989)}

\preref\ThreeJetsNLOME{R.K. Ellis, D.A. Ross and A.E. Terrano,
Phys.\ Rev.\ Lett. 45:1226 (1980); Nucl.\ Phys.\ B178:421 (1981)\semi
K. Fabricius, I. Schmitt, G. Kramer and G. Schierholz, 
Phys.\ Lett.\ B97:431 (1980); Z.\ Phys.\ C11:315 (1981)}

\preref\ThreeJetsPrograms{Z. Kunszt and P. Nason, in Z Physics at
LEP1, CERN Yellow Report 89-08\semi
G. Kramer and B. Lampe, Z. Phys.\ C34:497 (1987); C42:504(E) (1989);
Fortschr. Phys. 37:161 (1989)\semi
W.T. Giele and E.W.N. Glover, Phys.\ Rev.\ D46:1980 (1992)\semi
S. Catani and M.H. Seymour, Phys.\ Lett.\ B378:287 (1996),
hep-ph/9602277}

\preref\EventShapeAlphas{%
OPAL Collab., P.D. Acton et al., Z. Phys. C55:1 (1992)\semi
ALEPH Collab., D. Decamp et al., Phys. Lett. B284:163 (1992)\semi 
L3 Collab., O. Adriani et al., Phys. Lett. B284:471 (1992)\semi 
DELPHI Collab., P. Abreu et al., Z. Phys. C59:21 (1993)\semi
SLD Collab., K. Abe et al., Phys. Rev. D51:962 (1995)}

\preref\FourJetTests{S. Bethke, A. Ricker, P.M. Zerwas,
Z. Phys. C49:59-72 (1991);\semi
L3 Collab., B. Adeva et al., Phys. Lett. B248:227 (1990)\semi
DELPHI Collab., P. Abreu et al., Z. Phys. C59:357 (1993)\semi
OPAL Collab., R. Akers et al., Z. Phys. C65:367 (1995)}

\preref\PeskinSchroeder{%
M.E.\ Peskin and D.V.\ Schroeder, {\it An Introduction to Quantum Field Theory}
(Addison-Wesley, 1995)}

\preref\TwoLoopUnitarity{%
W.L.\ van Neerven, Nucl.\ Phys.\ B268:453 (1986)}

\preref\Adrian{A. Signer and L. Dixon, preprint hep-ph/9609460}

\preref\FiveGluon{%
Z. Bern, L. Dixon and D.A. Kosower, Phys.\ Rev. Lett.\
70:2677 (1993), hep-ph/9302280}

\preref\Kunsztqqqqg{%
Z. Kunszt, A. Signer and Z. Tr\'ocs\'anyi, Phys. Lett. B336:529 (1994),
hep-ph/9405386}

\preref\Color{%
F.A. Berends and W.T. Giele,
Nucl.\ Phys.\ B294:700 (1987)\semi
M.\ Mangano, S. Parke and Z.\ Xu, Nucl.\ Phys.\ B298:653 (1988)\semi
Z. Bern and D.A.\ Kosower, Nucl.\ Phys.\ B362:389 (1991)}

\preref\ManganoReview{%
M. Mangano and S.J. Parke, Phys.\ Rep.\ 200:301 (1991)\semi
L. Dixon, preprint hep-ph/9601359, to appear
in {\it Proceedings of TASI 95}, ed.\ D.E.\ Soper}

\preref\GG{W.T.\ Giele and E.W.N.\ Glover,
Phys.\ Rev.\ D46:1980 (1992)\semi
W.T.\ Giele, E.W.N.\ Glover and D. A. Kosower,
Nucl.\ Phys.\ B403:633 (1993)}

\preref\SpinorHelicity{%
F.A.\ Berends, R.\ Kleiss, P.\ De Causmaecker, R.\ Gastmans and T.\ T.\ Wu,
Phys.\ Lett.\ 103B:124 (1981)\semi
P.\ De Causmaeker, R.\ Gastmans,  W.\ Troost and  T.T.\ Wu,
Nucl. Phys. B206:53 (1982)\semi
R.\ Kleiss and W.J.\ Stirling, Nucl.\ Phys.\ B262:235 (1985)\semi
R.\ Gastmans and T.T.\ Wu,
{\it The Ubiquitous Photon: Helicity Method for QED and QCD}
(Clarendon Press, 1990)\semi
Z. Xu, D.-H.\ Zhang and L. Chang, Nucl.\ Phys.\ B291:392 (1987)}

\preref\GunionKunszt{%
J.F.\ Gunion and Z.\ Kunszt, Phys.\ Lett.\ 161B:333 (1985)}

\preref\StringBased{
Z. Bern and D.A.\ Kosower, \PRL 66:1669 (1991)\semi
Z. Bern and D.A.\ Kosower, \NPB 379:451 (1992)}

\preref\Mapping{Z. Bern and D.C.\ Dunbar,  Nucl.\ Phys.\ B379:562 (1992)}

\preref\Cutting{L.D.\ Landau, Nucl.\ Phys.\ 13:181 (1959)\semi
 S. Mandelstam, Phys.\ Rev.\ 112:1344 (1958), 115:1741 (1959)\semi
 R.E.\ Cutkosky, J.\ Math.\ Phys.\ 1:429 (1960)}

\preref\Massive{%
Z. Bern and A.G.\ Morgan, Nucl.\ Phys.\ B467:479 (1996), 
hep-ph/9511336}

\preref\KunsztFourPoint{%
Z. Kunszt, A. Signer and Z. Tr\'ocs\'anyi, Nucl.\ Phys.\ B411:397 
(1994), hep-ph/9305239\semi
A. Signer, Ph.D. thesis, ETH Z\"urich (1995)}

\preref\SusyDecomp{%
Z. Bern, hep-ph/9304249, in {\it Proceedings of Theoretical
Advanced Study Institute in High Energy Physics (TASI 92)},
eds.\ J. Harvey and J. Polchinski (World Scientific, 1993)\semi
Z.\ Bern and A.G.\ Morgan, Phys.\ Rev.\ D49:6155 (1994),
hep-ph/9312218\semi
A.G. Morgan, Phys.\ Lett.\ B351:249 (1995), hep-ph 9502230}

\preref\SusyFour{Z. Bern, L. Dixon, D.C. Dunbar and D.A. Kosower,
Nucl.\ Phys.\ B425:217 (1994), hep-ph/9405248}

\preref\SusyOne{Z. Bern, L. Dixon, D.C. Dunbar and D.A. Kosower,
Nucl.\ Phys.\ B435:39 (1995), hep-ph/9409265}

\preref\BDKconf{Z. Bern, L. Dixon and D.A. Kosower, hep-th/9311026,
in {\it Proceedings of Strings 1993}, eds. M.B. Halpern, A. Sevrin
and G. Rivlis (World Scientific, 1994)}

\preref\AllPlus{Z. Bern, G. Chalmers, L. Dixon and D.A. Kosower,
Phys.\ Rev.\ Lett.\ 72:2134 (1994), hep-ph/9312333}

\preref\Mahlon{G.D.\ Mahlon, Phys.\ Rev.\ D49:2197 (1994),
hep-ph/9311213; Phys.\ Rev.\ D49:4438 (1994), hep-ph/9312276}

\preref\Siegel{W. Siegel, Phys.\ Lett.\ 84B:193 (1979)\semi
D.M.\ Capper, D.R.T.\ Jones and P. van Nieuwenhuizen, Nucl.\ Phys.\
B167:479 (1980)\semi
L.V.\ Avdeev and A.A.\ Vladimirov, Nucl.\ Phys.\ B219:262 (1983)}

\preref\CollinsBook{J.C.\ Collins, {\it Renormalization}
(Cambridge University Press, 1984)}

\preref\IntegralsShort{Z. Bern, L. Dixon and D.A. Kosower,
Phys.\ Lett.\ 302B:299 (1993); erratum {\it ibid.} 318:649 (1993)}

\preref\IntegralsLong{Z. Bern, L. Dixon and D.A. Kosower,
\NPB 412:751 (1994)}

\preref\TreeCollinear{F.A. Berends and W.T. Giele, Nucl.\ Phys.\
B313:595 (1989)}

\preref\Factorization{%
Z. Bern and G. Chalmers, Nucl.\ Phys.\ B447:465 (1995), hep-ph/9503236}

\preref\Review{%
Z. Bern, L. Dixon and D.A. Kosower, to appear in
{\it Annual Reviews of Nuclear and Particle Science} (1996),
hep-ph/9602280}

\preref\Fermion{%
Z. Bern, L. Dixon and D.A. Kosower, Nucl.\ Phys. B437:259 (1995), 
hep-ph/9409393}

\preref\GloverMiller{E.W.N. Glover and D.J. Miller, preprint hep-ph/9609474}

\preref\Zqqgg{ 
Z. Bern, L. Dixon and D.A. Kosower, preprint hep-ph/9606378}

\preref\ZqqggFuture{
Z. Bern, L. Dixon and D.A. Kosower, in preparation}

\preref\IntegralsLong{Z. Bern, L. Dixon and D.A.\ Kosower,
\NPB 412:751 (1994), hep-ph/9306240}

\preref\IntegralsShort{Z. Bern, L. Dixon and D.A.\ Kosower,
Phys.\ Lett.\ 302B:299 (1993),
 erratum {\it ibid.} 318:649 (1993), hep-ph/9212308}

\preref\Lewin{L.\ Lewin, {\it Dilogarithms and Associated Functions\/}
(Macdonald, 1958)}

\preref\VNV{
W. van Neerven and J.A.M. Vermaseren, Phys.\ Lett.\ 137B:241 (1984)}

\preref\FiveGluon{Z. Bern, L. Dixon and D.A.\ Kosower, Phys.\ Rev.\ Lett.\
70:2677 (1993), hep-ph/9302280}

\preref\ThreeMassTriangle{H.-J. Lu and C. Perez, preprint
SLAC--PUB--5809 (1992)}

\preref\UDThreeMassTriangle{
N.I.\ Ussyukina and A.I.\ Davydychev, Phys.\ Lett.\ {298B:363 (1993)}} 
\preref\Gluinos{
S. Dawson, E. Eichten and C. Quigg, Phys.\ Rev.\ D31:1581 (1985)\semi
R.M. Barnett, H.E. Haber and G.L. Kane, Nucl. Phys. B267:625
(1986)\semi
L. Clavelli, Phys.\ Rev.\ D46:2112 (1992)\semi 
J. Ellis, D. Nanopoulos, and D. Ross, Phys.\ Lett.\ B305:375 (1993),
hep-ph/9303273\semi
L. Clavelli, P. Coulter and K. Yuan, Phys.\ Rev.\ D47:1973 (1993), 
hep-ph/9205237\semi
R. Mu\ttilde{n}oz-Tapia and W.J. Stirling, Phys.\ Rev.\ D49:3763 (1994),
hep-ph/9309246\semi
G.R. Farrar, Phys.\ Rev.\ D51:3904 (1995), hep-ph/9407401;
preprint hep-ph/9504295;
preprint hep-ph/9508291;
preprint hep-ph/9508292\semi
L. Clavelli, I. Terekhov, Phys.\ Rev.\ Lett.\ 77:1941 (1996), 
hep-ph/9605463; hep-ph/9603390\semi
A. de Gouvea and H. Murayama, preprint hep-ph/9606449\semi
Z. Bern, A.K.\ Grant, A.G.\ Morgan, preprint hep-ph/9606466}

\preref\HHKY{K. Hikasa, Mod. Phys. Lett. A5:1801 (1990)\semi
K. Hagiwara, T. Kuruma and Y. Yamada, Nucl. Phys. B358:80 (1991)}


\vskip -2. cm 
$\null$

\section{Introduction}
\tagsection\IntroSection

Electron-positron annihilation provides a clean experimental laboratory
for studying jet properties.  Leading-order predictions for the production
of up to five jets have been available for quite some time
[\use\ThreeJetsBorn,\use\FourJetsBorn,\use\ThreeJetsNLOME,\use\FiveJetsBorn],
but the reduction of theoretical uncertainties requires
next-to-leading-order (NLO) QCD corrections.  The NLO matrix elements for three-jet
production and other ${\cal O}(\alpha_s)$ observables are also 
known [\use\ThreeJetsNLOME], and numerical programs implementing 
these corrections [\use\ThreeJetsPrograms] have been widely used 
to extract a precise value of $\alpha_s$ from hadronic event 
shapes at the $Z$ pole~[\use\EventShapeAlphas].

Next-to-leading order corrections for more complicated processes are required,
however, if we wish
to use QCD in probing for new physics in other standard model processes.  
In $e^+e^-$ annihilation, for example, four-jet production is the lowest-order
process in which the quark and gluon color charges can be measured
independently.  Four-jet production is thus sensitive
to the presence of light colored fermions such as 
gluinos~[\use\Gluinos].
At LEP~2 the process $e^+e^- \to (\gamma^*,Z) \to
4$~jets is a background to threshold production of $W$ pairs, when
both $W$s decay hadronically.  The one-loop matrix elements required
for an NLO study of four-jet production are also needed for a
next-to-next-to-leading (NNLO) study of three-jet production at the
$Z$ pole.  Such a study (which awaits the computation of certain
two-loop matrix elements as well) would be desirable in order to 
reduce the theoretical uncertainties in determining $\alpha_s$
via this process.

In this paper we present analytic formulas for the one-loop helicity
amplitudes for electron-positron annihilation into four-quarks,
$e^+e^- \to (\gamma^*,Z) \to \qb \q \Qb \Q$.  Together with the
leading-in-color matrix elements for production of two quarks and two
gluons, $e^+e^- \rightarrow \qb \q g g$ [\use\Zqqgg,\use\ZqqggFuture],
the leading-color parts of these amplitudes have already been
incorporated into an NLO program for $e^+e^- \rightarrow 4$~jets
[\use\Adrian].  The same amplitudes presented here may also be used in
computations of $W$ or $Z + 2$ jet production at hadron colliders and
three-jet production in deeply inelastic scattering.  Glover and
Miller~[\use\GloverMiller] have also recently reported on a calculation of
the squared one-loop matrix elements for $e^+e^- \to \gamma^* \to \qb
\q \Qb \Q$, summed over helicities and expressed in terms of Lorentz
scalar products of the quark four-momenta, rather than the spinor
products that we employ.  A comparison of
their results with those presented in this paper will be useful.

Recent years have seen a number of technical advances in the
computation of one-loop amplitudes, which the authors have surveyed
in a recent review article~[\use\Review].  
These advances have made possible the calculation of all
one-loop five-parton processes
[\use\FiveGluon,\Kunsztqqqqg,\use\Fermion], as well as of a number of infinite
sequences of one-loop amplitudes~[\use\AllPlus,\use\Mahlon,\use\SusyFour,\use\SusyOne].
  The general
strategy employed in this paper (and in a subsequent paper on $e^+e^-
\rightarrow \qb \q gg$ [\use\Zqqgg,\use\ZqqggFuture]) is to obtain
amplitudes from their analytic structure.  In particular, we use the
constraints of
unitarity~[\use\Cutting,\use\SusyFour,\use\SusyOne,\use\Massive] and
factorization [\use\AllPlus,\use\Factorization], as summarized in
ref.~[\use\Review].  This approach leads to relatively compact
expressions, as compared with those obtained from 
a traditional, diagrammatic computation.  In this
approach, amplitudes obtained previously are recycled to obtain new
amplitudes; manifest gauge invariance is therefore maintained.  In a
Feynman diagram approach each diagram alone is not gauge
invariant, and are often individually 
much more complicated than a final sum over
diagrams.

We use helicity methods [\use\SpinorHelicity,\use\GunionKunszt] 
since they lead to relatively compact expressions for the amplitudes, 
and retain all spin information.
We also make use of color decompositions [\use\Color] to
help simplify the analytic structures that must be computed.
As a check, we have verified numerically that the amplitudes presented 
in this paper agree with a direct Feynman diagram calculation 
we performed.

The paper is organized as follows. In section~\use\BasicToolsSection,
we briefly describe helicity methods and color decompositions.  
We give the
amplitudes, together with a brief description of the calculational 
methods, in section~\use\AmplitudeSection; we describe the contribution
proportional to the axial vector coupling of the $Z$ to the $t,b$ quark
isodoublet in subsection~\use\AxialSection. 
A summary is included in section~\use\SummarySection.  
We collect descriptions of various integral functions appearing 
in the amplitudes in an appendix.

\section{Basic Tools}
\tagsection\BasicToolsSection

In this section, we briefly review two of the basic tools useful for expressing
amplitudes in a compact form: the spinor helicity method and color
decompositions.  The reader is referred to review articles
[\use\ManganoReview] and references therein for further details.

\subsection{Spinor Helicity}

In explicit calculations it is usually convenient to use a helicity
basis, where all quantities are rewritten in terms of Weyl spinors
$\vert k^{\pm} \rangle$.  
Although there are no external gluons in the final 
$e^+e^-\to \qb \q\Qb \Q$ helicity amplitudes, they appear as intermediate
states in various unitarity cuts and factorization limits that are used 
to construct the amplitudes.  
We made use of the gluon polarization vectors of
Xu, Zhang and Chang [\use\SpinorHelicity,\use\GunionKunszt], 
$$
\pol^{+}_\mu (k;q) = 
     {\sandmm{q}.{\gamma_\mu}.k
      \over \sqrt2 \spa{q}.k}\, ,\hskip 1cm
\pol^{-}_\mu (k;q) = 
     {\sandpp{q}.{\gamma_\mu}.k
      \over \sqrt{2} \spb{k}.q} \, ,
\eqn\HelicityDef
$$  
where $k$ is the gluon momentum and $q$ is an arbitrary null
`reference momentum' which drops out of final gauge-invariant
amplitudes.  
The plus and minus labels on the polarization vectors
refer to the gluon helicities.  
Our (crossing-symmetric) convention takes all particles to be
outgoing, and labels the helicity 
and particle vs. antiparticle assignment accordingly.
(That is, we write the amplitudes for the process
$0\rightarrow V \qb \q \Qb\Q$.)
For incoming (negative energy) momenta the helicity 
and particle vs. antiparticle assignment are reversed.
It is convenient to define 
$$
\eqalign{
\spa{i}.j &\equiv  \langle k_i^{-} \vert k_j^{+} \rangle\,, 
\hskip 2.0 cm 
\spb{i}.j \equiv \langle k_i^{+} \vert k_j^{-} \rangle \,, \cr
\spab{i}.{l}.{j} &\equiv 
\langle k_i^- \, |\, \ksl_l\, | \, k_j^- \rangle \,, 
\hskip 1.0 cm
\spab{i}.{(l+m)}.{j} \equiv 
\langle k_i^- \, |\, (\ksl_l+\ksl_m) \, | \, k_j^- \rangle \,, \cr
&\hskip -5mm \spaa{i}.{(l+m)(n+r)\cdots}.{j} \equiv 
\langle k_i^- \, |\, (\ksl_l+\ksl_m)(\ksl_n+\ksl_r)\cdots \, | \, k_j^+ \rangle \,, \cr
}\eqn\AngleDef
$$ 
for the spinor products, which is the notation 
we shall use to quote the results. 
Here all the momenta $k_i$ are massless.
The spinor inner products $\spa{i}.j$, $\spb{i}.j$ are antisymmetric 
and satisfy $\spa{i}.j \spb{j}.i = 2 k_i \cdot k_j \equiv s_{ij}$.

To maximize the benefit obtained from the spinor helicity formalism for loop
amplitudes we must choose a compatible regularization scheme.  In
conventional dimensional regularization [\use\CollinsBook], the
polarization vectors are $(4-2\eps)$-dimensional, which is
incompatible with the spinor helicity method's use of
four-dimensional polarizations.  
To avoid this problem, we modify the regularization scheme so all 
helicity states are four-dimensional and
only the loop momentum is continued to $(4-2\eps)$ dimensions.  
This is the four-dimensional-helicity (FDH) scheme~[\use\StringBased], 
which has been shown to be equivalent at one-loop~[\use\KunsztFourPoint] to an 
appropriate helicity formulation of Siegel's dimensional-reduction 
scheme (\DRbar)~[\use\Siegel].  
The conversion between schemes has been given in 
ref.~[\use\KunsztFourPoint]; there is no loss of generality 
in choosing the FDH/\DRbar\ scheme.

\subsection{Color Decomposition}

In this section we describe a color decomposition [\use\Color] of the
one-loop amplitude for $e^+ e^- \rightarrow \qb \q \Qb \Q$, in terms
of group-theoretic factors (color structures) multiplied by kinematic
functions called {\it partial amplitudes}.  Because of the crossing
symmetry of the spinor products [\use\GunionKunszt] and of the
integral functions (discussed in the appendix), these partial
amplitudes may also be used to obtain the one-loop contributions to
$Z$ or $W + 2$ jet production at hadron colliders or three jet
production in deeply inelastic scattering.  (For the case of the $W$,
only the coupling constants need be changed in the formul\ae\ given
below.)

The partial amplitudes are defined to be the coefficients of
the various color structures.  Consider the amplitude 
$\A{6}(1_\q,2_\Qb,3_\Q,4_\qb;5_\eb,6_e)$. 
At tree-level its decomposition is 
$$
\eqalign{
\A{6}^{\rm tree} (1_\q,2_\Qb,3_\Q,4_\qb) & = 
 2 e^2 g^2 \biggl[  \Bigl( - Q^\q 
+  v_{L,R}^e v_{L,R}^\q \,\prop{Z}(s_{56}) \Bigr) 
  A_{6}^{\rm tree}(1_\q,2_\Qb,3_\Q,4_\qb) \cr
& \hskip 1.5 cm 
+  \Bigl( - Q^\Q 
+ v_{L,R}^e v_{L,R}^\Q \,\prop{Z}(s_{56})\Bigr) 
  A_{6}^{\rm tree}(3_\Q,4_\qb,1_\q,2_\Qb) \biggr] \cr
& \hskip 1 cm \times 
\Bigl(\delta_{i_1}^{\ib_2} \,  \delta_{i_3}^{\ib_4}  \, 
-{1\over N_c} \delta_{i_1}^{\ib_4}\,  \delta_{i_3}^{\ib_2} \,\Bigr) \,,\cr}
\eqn\TreeColorDecomposition
$$
where we have suppressed the $5,6$ labels of the electron pair, $e$ is
the QED coupling, $g$ the QCD coupling, $Q^\q$ ($Q^\Q$) is the charge of 
quark $q$ (quark $Q$) in units of $e$, 
and the left- and right-handed couplings are
$$
\eqalign{
v_L^e & = { -1 + 2\sin^2 \theta_W \over \sin 2 \theta_W } \,, 
\hskip 2.3 cm 
v_R^e  = { 2 \sin^2 \theta_W \over  \sin 2 \theta_W } \,,  \cr 
v_L^q & = { \pm 1 - 2 Q^q\sin^2 \theta_W \over  \sin 2 \theta_W } \,,
\hskip 1.9 cm 
v_R^q = -{ 2 Q^q \sin^2 \theta_W \over \sin 2 \theta_W }  \,, \cr }
\anoneqn
$$ 
where $\theta_W$ is the Weinberg angle.
Eq.~(\use\TreeColorDecomposition) contains the ratio of the $Z$ and
photon propagators,
$$\prop{Z}(s) = {s \over s - M_Z^2 + i \Gamma_Z \, M_Z}
\anoneqn$$
where $M_Z$ and $\Gamma_Z$ are the mass and width of the $Z$.  
We have given the decomposition for a general $SU(N_c)$ gauge group;
$N_c=3$ for QCD.
The two signs in $v_L^q$ correspond to up $(+)$ and down $(-)$ type quarks. 

The subscripts $L$ and $R$ refer to whether the particle to which 
the $Z$ couples is left- or right-handed. 
That is, $v_R^q$ is to be used for the configuration
where the quark (leg 1) has plus helicity and the anti-quark (leg 4) 
has minus helicity, which we denote by the shorthand $(1_\q^+, 4_\qb^-)$.
Similarly, $v_L^q$ corresponds to the configuration  $(1_\q^-, 4_\qb^+)$.
Because the electron and positron are incoming in
$e^+ e^-$ annihilation, our outgoing-momenta notation reverses 
their helicities and particle vs. anti-particle assignment.
Thus, $v_R^e$ corresponds to the helicity configuration $(5_\eb^-, 6_e^+)$
whereas $v_L^e$ corresponds to the configuration $(5_\eb^+, 6_e^-)$.

We have defined the tree and one-loop partial amplitudes $A_6^\tree$ and
$A_{6;i}$ to include a photon propagator.  
The ratio $\prop{Z}(s_{56})$ appearing in 
eq.~(\use\TreeColorDecomposition) then replaces the photon propagator with
a $Z$ propagator.  
This form of the amplitude is convenient for checking
that amplitudes properly reduce to lower-point amplitudes when the
$e^+$ and $e^-$ momenta are taken to be collinear.

At one loop there are three partial amplitudes,
$$
\hskip -5mm\eqalign{
\A{6}^{1\rm -loop} &(1_\q ,2_\Qb,3_\Q,4_\qb)  =  \cr
&\hskip -5mm 2 e^2 g^4  \biggl[\Bigl( - Q^\q 
+  v_{L,R}^e v_{L,R}^\q \prop{Z}(s_{56}) \Bigr) 
%
\Bigl[ N_c \,  \delta_{i_1}^{\ib_2} \, \delta_{i_3}^{\ib_4}  \, 
         A_{6;1}(1_\q,2_\Qb,3_\Q,4_\qb)
   + \delta_{i_1}^{\ib_4}\,  \delta_{i_3}^{\ib_2} \,
           A_{6;2} (1_\q,2_\Qb,3_\Q,4_\qb) \Bigr] \cr
&
+ \Bigl( - Q^\Q 
+  v_{L,R}^e v_{L,R}^\Q \,\prop{Z}(s_{56}) \Bigr) 
%
\Bigl[N_c \,  \delta_{i_1}^{\ib_2} \, \delta_{i_3}^{\ib_4}  \, 
         A_{6;1}(3_\Q,4_\qb,1_\q,2_\Qb)
   + \delta_{i_1}^{\ib_4}\,  \delta_{i_3}^{\ib_2} \,
           A_{6;2} (3_\Q,4_\qb,1_\q,2_\Qb) \Bigr] \cr
&
+ { v_{L,R}^e \over \sin2\theta_W }\prop{Z}(s_{56}) \,
%
\Bigl( \delta_{i_1}^{\ib_2} \, \delta_{i_3}^{\ib_4}  
     - {1\over N_c} \delta_{i_1}^{\ib_4}\,  \delta_{i_3}^{\ib_2} \Bigr)
      A_{6;3} (1_\q,2_\Qb,3_\Q,4_\qb) \biggr] \,.\cr
}\eqn\LoopColorDecomposition
$$
Note the additional factor of $N_c$ in the color-tensor coefficients
of $A_{6;1,2}$ compared to the corresponding tensors in the tree-level
amplitude.  The $A_{6;3}$ term arises from a fermion triangle
graph~[\use\HHKY].  It violates the axial symmetry, and is
proportional to the axial coupling of the $Z$ to the top and bottom
quark isodoublet as well as to the top--bottom mass splitting (see
section~\use\AxialSection).  The corresponding charges vanish for the
photon and the $W$ boson, and this term does not contribute to
amplitudes for the latter bosons.

Helicity conservation for massless quarks and leptons ensures that
there are only $2^3=8$ possible helicity configurations for 
$e^+e^- \to \qb \q \Qb \Q$, namely two choices for each fermion line.  
Because the electron line couples through the current 
$\langle5^\pm | \gamma^\mu | 6^\pm \rangle 
= \langle6^\mp | \gamma^\mu | 5^\mp \rangle$,
it is trivial to reverse its helicity simply by exchanging
$5$ and $6$ in the partial amplitudes $A_6^\tree$ and $A_{6;i}$,
and exchanging $v_R^e \lr v_L^e$ in the prefactors in 
eqs.~(\use\TreeColorDecomposition) and~(\use\LoopColorDecomposition)
as discussed above.
We can use parity to reverse all helicities simultaneously
in the partial amplitudes $A_6^\tree$, $A_{6;1}$ and $A_{6;2}$, 
by complex conjugating all spinor products ($\spa{i}.j \lr \spb{j}.i$).
The axial vector contribution $A_{6;3}$ is also complex conjugated, 
but acquires an additional overall minus sign from the $\gamma_5$ 
in the loop.
Thus we are left with just two independent helicity configurations,
which we may take to be
$\A{6}(1_\q^+,2_\Qb^\pm,3_\Q^\mp,4_\qb^-;5_\eb^-,6_e^+)$
(we shall again suppress the $5,6$ labels below).

The partial amplitudes can be further expressed in terms of `primitive
amplitudes' [\use\Fermion].  The primitive amplitudes are
gauge-invariant classes of color-stripped amplitudes from which we can
build the partial amplitudes.  (In ref.~[\use\Fermion] the primitive
amplitudes were defined to have a fixed ordering of the external legs, 
but here we extend the notion to mean gauge-invariant color-stripped 
building-blocks for amplitudes.)  
It turns out that some signs in the reduction of partial amplitudes to
primitive amplitudes depend on which of the two helicity configurations
one is considering, so for clarity we shall explicitly list the two 
cases separately.  
Although color decompositions do not depend on 
the helicity choices, these sign differences appear because we 
have used the symmetries of the primitive amplitudes to reduce the number 
of independent ones required.  

The formul\ae\ for $A_{6;i}(1_\q^+,2_\Qb^+,3_\Q^-,4_\qb^-)$
in terms of the primitive amplitudes are
$$
\eqalign{
 A_{6;1}(1_\q^+,2_\Qb^+,3_\Q^-,4_\qb^-) &
=  A^\nn_6(1,2,3,4)
+ {n_s - n_{\!f} \over N_c} A^{s\,,\nn}_6(1,2,3,4)
- {n_{\! f} \over N_c} A^{\!f,\,\nn}_6(1,2,3,4) \cr
& \hskip .5 cm 
- {2\over N_c^2} \bigl(A^\nn_6(1,2,3,4) +
                       A^\an_6(1,3,2,4) \bigr) 
+ {1\over N_c^2} A_6^{\sl}(2,3,1,4) \,,\cr\cr
A_{6;2}(1_\q^+,2_\Qb^+,3_\Q^-,4_\qb^-) &= 
A^\an_6(1,3,2,4) 
- {n_s - n_{\!f} \over N_c} A^{s,\,\nn}_6(1,2,3,4)
 + {n_{\! f} \over N_c} A^{\!f,\,\nn}_6(1,2,3,4) \cr
& \hskip .5 cm 
+ {1\over N_c^2} \bigl(A^\an_6(1,3,2,4) +
                       A^\nn_6(1,2,3,4) \bigr)
- {1\over N_c^2} A_6^{\sl}(2,3,1,4) \,, \cr\cr
A_{6;3}(1_\q^+,2_\Qb^+,3_\Q^-,4_\qb^-) &= A^\ax_6(1,4,2,3) \,,\cr
}\eqn\nndecomp
$$
where $n_s$ is the number of scalars%
\footnote{$^*$}{As in refs.~[\use\FiveGluon,\use\Fermion], each 
scalar here contains four states (to match the four states of Dirac
fermions) so that $n_s$ must be divided by two for comparisons to
conventional normalizations of scalars.}  ($n_s = 0$ in QCD) and
$n_{\! f}$ is the number of Dirac fermions.  Within the context of
supersymmetry decompositions
[\use\FiveGluon,\use\SusyDecomp,\use\Review] it is natural to divide
the fermion contribution into $s$ and $f$ pieces as we have done here.

For the other independent helicity configuration 
$A_{6;i}(1_\q^+,2_\Qb^-,3_\Q^+,4_\qb^-)$, 
we have, 
$$
\eqalign{
 A_{6;1}(1_\q^+,2_\Qb^-,3_\Q^+,4_\qb^-) &
=  A_6^\an(1,2,3,4)
 + {n_s - n_{\!f} \over N_c} A^{s,\,\an}_6(1,2,3,4)
- {n_{\! f} \over N_c} A^{\!f,\,\an}_6(1,2,3,4) \cr
& \hskip .5 cm 
- {2 \over N_c^2}\bigl( A_6^\an(1,2,3,4) 
                      + A_6^\nn(1,3,2,4) \bigr)
- {1\over N_c^2}  A_6^{\sl}(3,2,1,4) \,,\cr\cr
A_{6;2}(1_\q^+,2_\Qb^-,3_\Q^+,4_\qb^-) &= 
 A_6^\nn(1,3,2,4)
- {n_s - n_{\!f} \over N_c} A^{s,\,\an}_6(1,2,3,4)
 + {n_{\! f} \over N_c} A^{\!f,\,\an}_6(1,2,3,4) \cr
& \hskip .5 cm 
+ {1\over N_c^2} \bigl(  A^\nn_6(1,3,2,4) 
                       + A^\an_6(1,2,3,4) \bigr)
+ {1\over N_c^2}  A_6^{\sl}(3,2,1,4) \,,\cr\cr
A_{6;3}(1_\q^+,2_\Qb^-,3_\Q^+,4_\qb^-) &= -A^\ax_6(1,4,3,2) \,.\cr
}\eqn\andecomp
$$
Note that $A_6^\sl$ and $A_6^\ax$ appear with a different permutation of the
arguments,
and the opposite sign, compared with the previous helicity structure.  
The signs in the above expressions also depend on the relative phase
conventions of the two helicity amplitudes; we have chosen a convention
where $A_6^{\tree,\an}(1,3,2,4)=-A_6^{\tree,\nn}(1,2,3,4)$.  Of course,
consistency between tree and loop amplitudes must be maintained.

A representative diagram for the primitive amplitudes 
$A_6^{s,\bn}(1,2,3,4)$ and $A_6^{f,\bn}(1,2,3,4)$, 
corresponding to those terms 
in the partial amplitudes proportional to the number of fermions 
$n_{\!f }$, is depicted in \fig\FermionBubbleFigure. 
There are only two such diagrams.  
The `parent' pentagon diagram for the leading-color helicity 
amplitude $A_6^\bn(1,2,3,4)$ is depicted in \fig\LeadingPentFigure. 
(By a `parent' diagram we mean a diagram from which all other diagrams 
in the set can be obtained via a continuous `pinching' process,
in which two lines attached to the loop are brought together to a
four-point interaction --- if such an interaction exists --- 
or further pulled out from the loop, and left
as the branches of a tree attached to the loop.)  
The parent box and triangle diagrams for the 
subleading-color 
primitive amplitude $A_6^\sl(1,2,3,4)$ are depicted in \fig\SubleadingSumFigure.
(We combine the triangle-parent contributions with the box-parent
contributions because the former are so simple, and the two have an
identical color structure.)  The fermion-loop triangle diagram,
proportional to the axial coupling of the $Z$ to quarks, is shown in
\fig\AxialTriFigure.  We neglect the $u,d,s,c$ quark masses, and so only the
$t,b$ quark pair survives an isodoublet cancellation in the loop,
thanks to its large mass splitting.

\LoadFigure\FermionBubbleFigure{\baselineskip13pt\narrower\ninerm 
One of the two fermion bubble diagrams contributing to 
$A_6^{s,\bn}$ and $A_6^{f,\bn}$.}
{\epsfysize 2.0truein}
{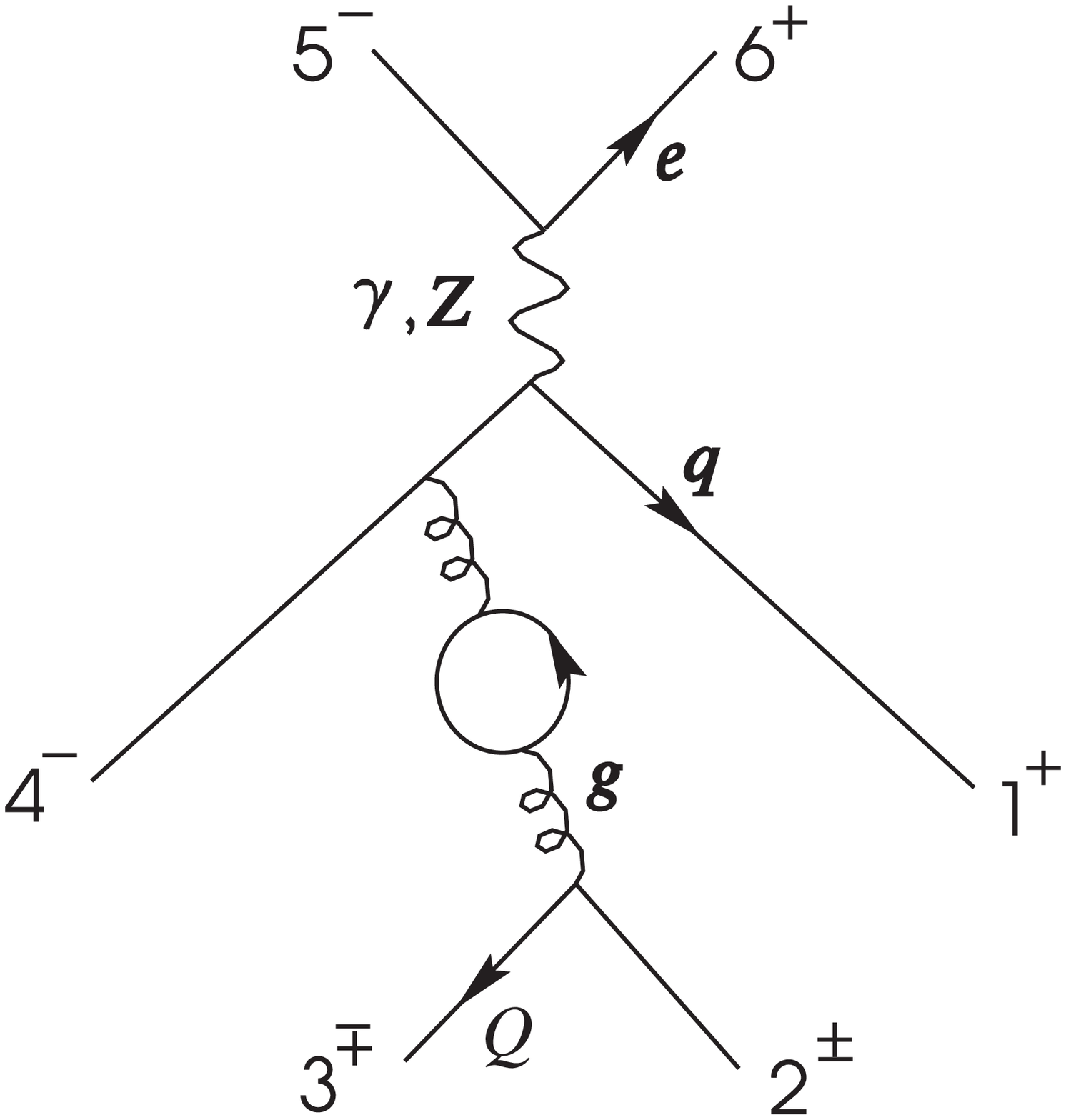}

\LoadFigure\LeadingPentFigure{\baselineskip13pt\narrower\ninerm 
The parent diagram for the leading-color helicity amplitudes 
$A_6^\bn$.}
{\epsfysize 2.0truein}
{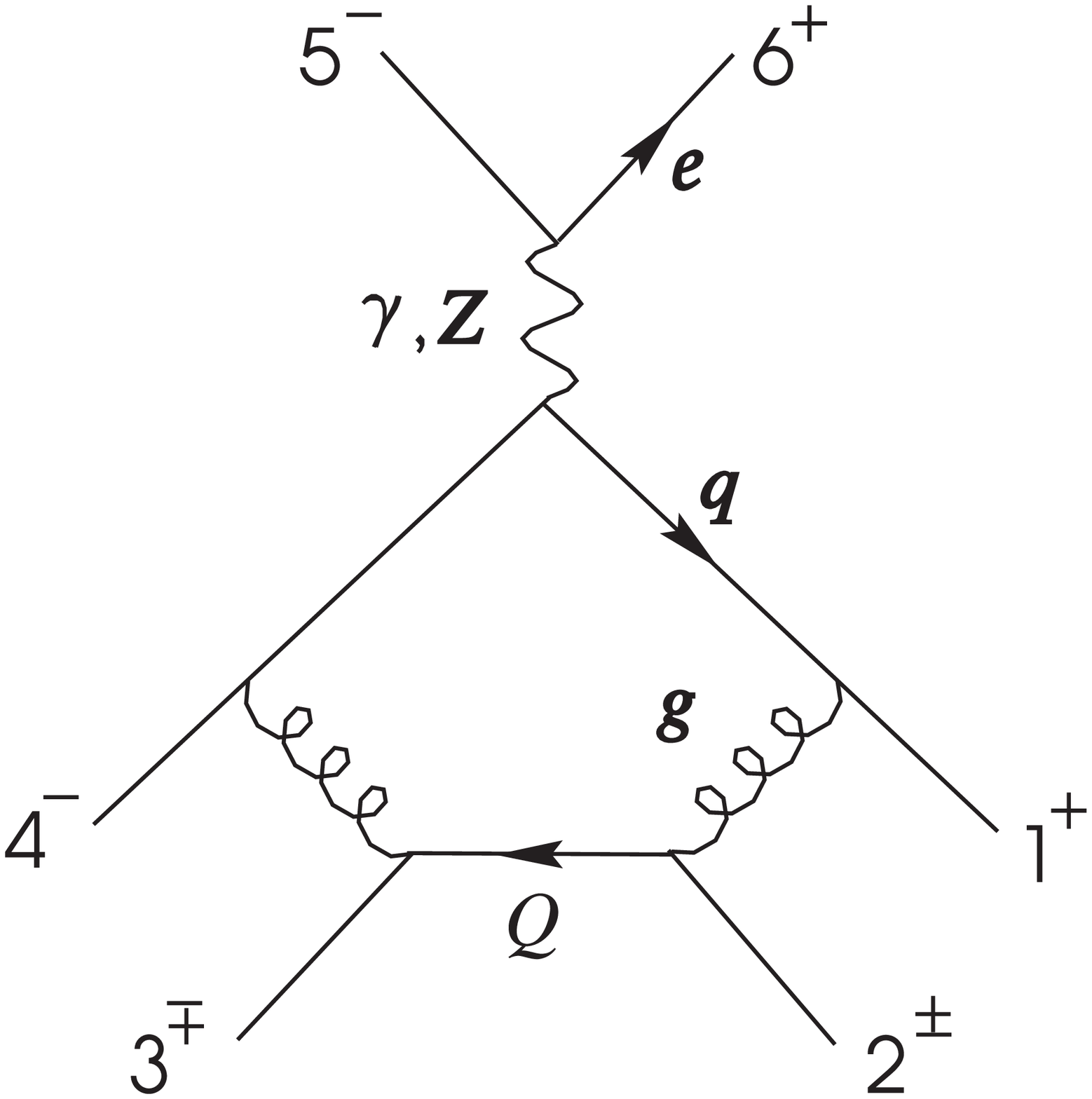}

\LoadFigure\SubleadingSumFigure{\baselineskip13pt\narrower\ninerm 
The diagrams for the subleading-color component 
$A_6^\sl$ can be divided into (a) those with a
parent box diagram), and 
(b) those with a parent triangle diagram. Note that the labelling
of legs is different from that in the leading-color diagrams.} 
{\epsfysize 2.3truein}{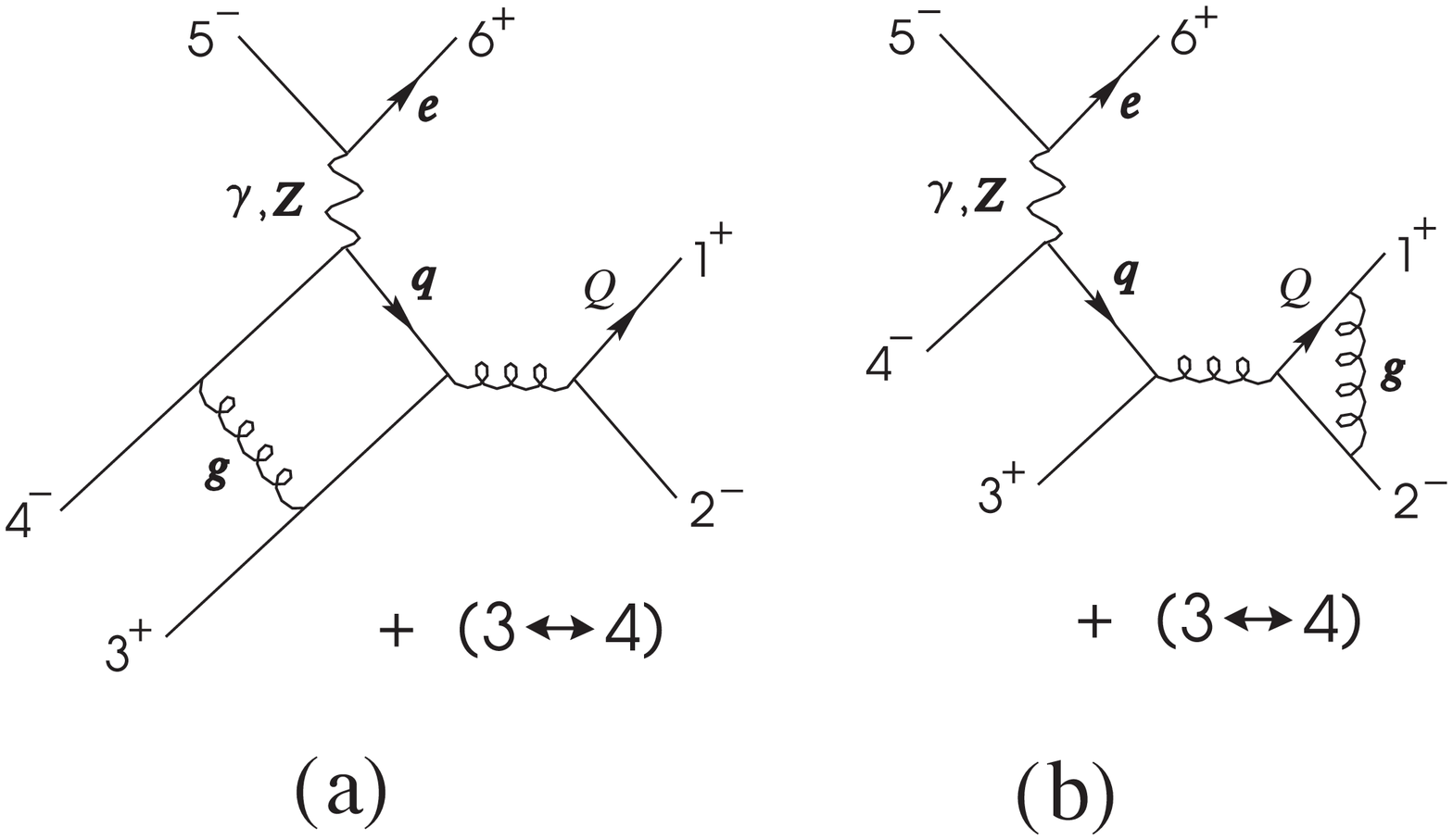}

\LoadFigure\AxialTriFigure{\baselineskip13pt\narrower\ninerm 
Triangle diagram for the amplitude component $A_6^\ax$, 
proportional to the axial coupling of the $Z$ to quarks.} 
{\epsfysize 2.5truein}{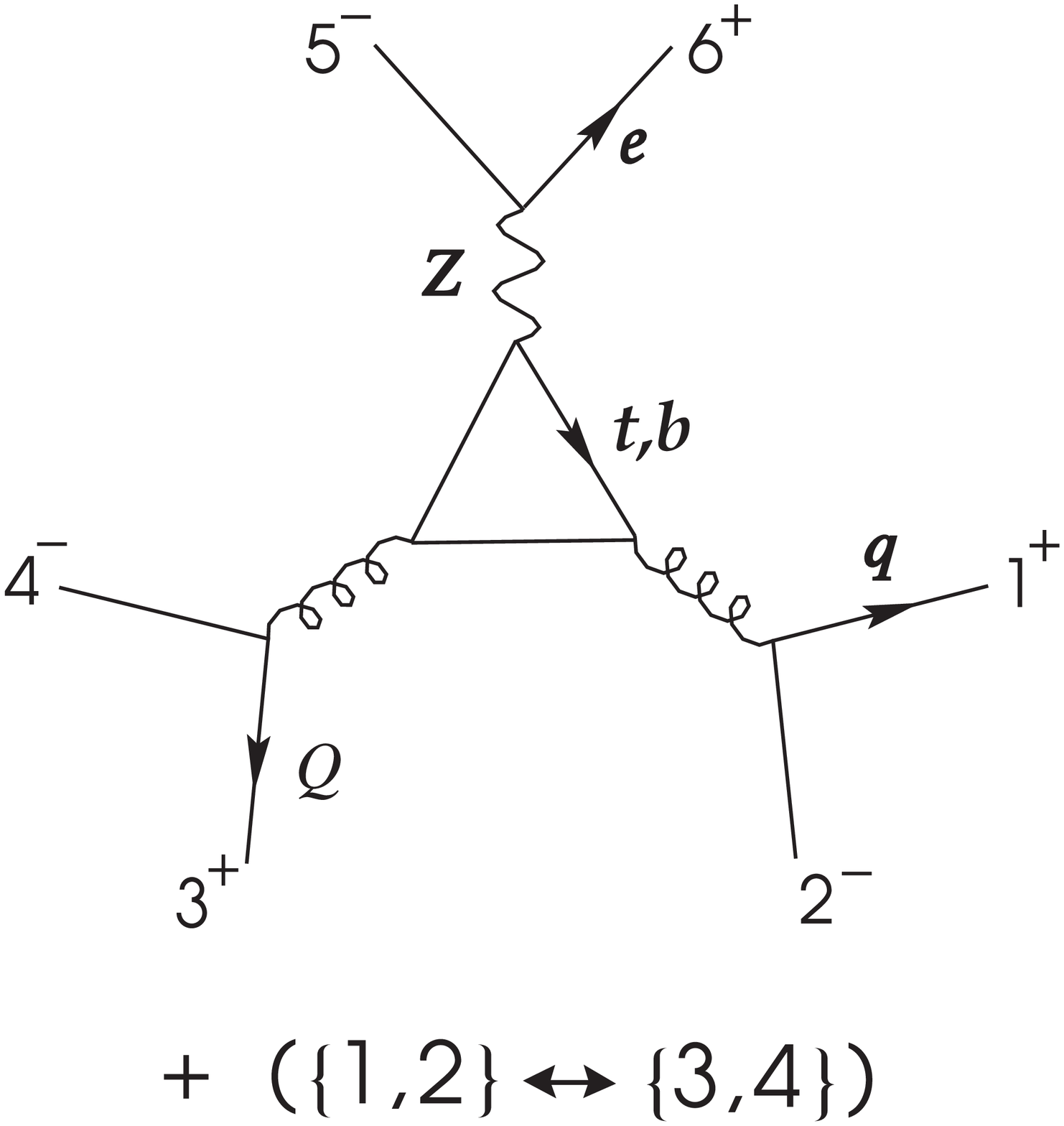}

Formulas~(\use\TreeColorDecomposition) and
(\use\LoopColorDecomposition) apply to the case of unequal quark
flavors, $q\neq Q$.  The equal flavor amplitude may be obtained
from the unequal-flavor formula by subtracting the same formula with
$q$ and $Q$ exchanged ({\it or} $\qb$ and $\Qb$ exchanged,
but not both),
and then setting $Q=q$ in all the coupling constant prefactors.
(Equal-flavor cross sections also require
 an identical-particle factor in the phase-space 
measure, which is $(\hf)^2$ for $e^+e^-\to\qb q \qb q$.)

The virtual part of the next-to-leading order correction to the
parton-level cross-section is given by
$$
d\sigma_6^{{\rm NLO,virtual}} 
= 2 \sum_{{\rm colors}} \Re \bigl[ 
 \A{6}^{\tree\,*} \, \A{6}^{1-{\rm loop}} \bigr] \, .
\anoneqn
$$
It is a straightforward exercise (which we leave to the reader) 
to evaluate this color sum in terms of partial amplitudes.

\section{The Amplitudes}
\tagsection\AmplitudeSection

In constructing the amplitudes presented here, we use two basic
analytic properties: that their imaginary (absorptive) parts be
determined from the Cutkosky rules [\use\Cutting], 
and that they factorize on particle poles.  
These analytic properties of amplitudes have, of course, played an 
important role in field theory for many decades; 
we make use of recent developments which allow us to to obtain 
complete amplitudes with no subtractions.  
In order to maximize the efficiency
of the computation it is useful to perform the computation in the
manner described in ref.~[\use\Review].  Due to the complexity of the
kinematics for the processes presented here, further techniques are
required to minimize the appearance of undesired spurious poles; these
will be discussed in ref.~[\use\ZqqggFuture].

The primitive amplitudes $ A^{\! f}_6$ and $A^s_6$ are proportional to
tree amplitudes and are given by
$$
\eqalign{
A_6^{s,\bn}(1,2,3,4)
 &= \cg A_6^\tree(1_\q^+,2_\Qb^\pm,3_\Q^\mp,4_\qb;5_\eb^-,6_e^+)
  \left[ -{1\over 3\eps} \L {\mu^2 \over -s_{23}}\R^{\eps} 
          - {8\over 9} \right] \,, \cr
A_6^{\! f,\bn}(1,2,3,4)
 &= \cg A_6^\tree(1_\q^+,2_\Qb^\pm,3_\Q^\mp,4_\qb;5_\eb^-,6_e^+)
  \left[ {1\over\eps} \L {\mu^2 \over -s_{23}}\R^{\eps} 
          + 2 \right] \,, \cr}
\eqn\Afsdef
$$   
where $A_6^{\rm tree}$ is given below,
$$\eqalign{
  \cg\ &=\ {1\over (4\pi)^{2-\eps}}
 {\Gamma(1+\eps)\Gamma^2(1-\eps)\over\Gamma(1-2\eps)}\,, \cr
}\eqn\cgammadef
$$
and $D= 4 - 2 \eps$.

The contribution $A_6^\ax$ is finite (see below).
It is convenient to decompose the remaining primitive amplitudes into
divergent ($V$) and finite ($F$) pieces,
$$
A^{\alpha}_6\ =\ \cg \Bigl[ 
A^{{\rm tree},\, \alpha }_6 V^\alpha + i \, F^\alpha \Bigr]\, ,
\eqn\VFdecomp
$$
where 
$$
\alpha = \{ \nn\,, \quad \an\,, \quad \sl \}
\anoneqn
$$
labels the primitive amplitude under discussion. 
The quantities $A_6^{\tree,\bn}$ coincide with the true tree partial
amplitudes appearing in eq.~(\use\TreeColorDecomposition),
$$
A_6^\tree(1_\q^+,2_\Qb^\pm,3_\Q^\mp,4_\qb;5_\eb^-,6_e^+) 
 = A_6^{\tree,\bn}(1,2,3,4)\,.
\anoneqn  
$$
(We continue to omit the arguments of the primitive amplitudes
corresponding to the lepton pair.)
We express the amplitudes in terms of the spinor
products~(\use\AngleDef), the Lorentz products
$s_{ij} = (k_i+k_j)^2$ and $t_{ijm} = (k_i+k_j+k_m)^2$, 
and the following combinations of kinematic invariants,
$$ 
\eqalign{
& \d12  = s_{12} - s_{34} - s_{56}\, , \hskip 1.3 cm 
\d34  = s_{34} - s_{56} - s_{12}\, , \hskip 1.3 cm 
\d56  = s_{56} - s_{12} - s_{34}\, , \cr
& \hskip 2.8 cm 
\dt = s_{12}^2 + s_{34}^2 + s_{56}^2 - 2 s_{12}s_{34} - 2 s_{34} s_{56}
 - 2 s_{56} s_{12} \,. \cr}
\eqn\deltaijdef
$$
The latter quantity is the negative of the 
Gram determinant associated with the set of massive
momenta $\{k_1+k_2,k_3+k_4,k_5+k_6\}$.

The amplitudes we present are bare ones, i.e., no ultra-violet subtraction
has been performed.  To obtain the renormalized amplitudes in 
an \MSbar -type subtraction scheme, one should subtract the quantity
$$
c_\Gamma N_c \, g^2 \left[ {1\over\e}\left( {11\over3}
  - {2\over3}{\nf\over N_c} - {1\over3}{\ns\over N_c} \right) \right]
    \A{6}^\tree\,,
\eqn\mssubtraction
$$
from the the amplitude (\use\LoopColorDecomposition). 

We quote the results in the FDH scheme [\use\Siegel,\use\StringBased],
but these may easily be converted to the 't~Hooft-Veltman scheme;
to do so one would add the quantity  
$$
- \cg N_c \, g^2 \L {{2\over 3} - {1\over N_c^2}} \R \A6^\tree 
\eqn\schemeshift
$$
to the amplitude (\use\LoopColorDecomposition) and change the
coupling constant from the non-standard $\alpha_{\overline{DR}}$ to
the standard $\alpha_{\overline{MS}}$.  The conversion between the various
schemes is discussed in refs.~[\use\KunsztFourPoint].

\subsection{The Helicity Configuration $q^+\Qb^+Q^-\qb^-$ }

We first give the primitive amplitude $A_6^\nn(1,2,3,4)$ which contributes
to the leading color part of 
$A_{6;1}(1_\q^+,2_\Qb^+,3_\Q^-,4_\qb^-;5_\eb^-,6_e^+)$.
This amplitude is odd under a `flip symmetry', which is the combined
operation of a permutation and spinor-product complex conjugation:
$$
\hbox{flip:} \hskip 8 mm
1\leftrightarrow 4\,, \hskip 0.3 cm  
2\leftrightarrow 3\,, \hskip 0.3 cm  
5\leftrightarrow 6\,, \hskip 0.3 cm 
\spa{a}.{b} \leftrightarrow \spb{a}.{b} \,, \hskip 0.3 cm 
\sandmm{a}.{b}.{c} \leftrightarrow \sandpp{a}.{b}.{c} = \sandmm{c}.{b}.{a} \,.
\eqn\FlipSymmetry
$$
The corresponding tree amplitude for this helicity configuration is
$$
A^{{\rm tree}, \, \nn}_6(1,2,3,4) =
  i \, {\spb1.2\spa5.4\spab3.{(1+2)}.6 \over s_{23} s_{56} t_{123}}
+ i \, {\spa3.4\spb6.1\spab5.{(3+4)}.2 \over s_{23} s_{56} t_{234}} \, .
\anoneqn
$$
We have
$$
\eqalign{
V^\nn(1,2,3,4) =  
-{1\over \eps^2} \L \L{\mu^2 \over -s_{12}}\R^{\eps}
 + \L {\mu^2 \over -s_{34}}\R^{\eps} \R
 + {2\over 3\eps} \L{\mu^2 \over -s_{23}} \R^{\eps} 
 - {3\over 2} \ln\Bigl({-s_{23}\over -s_{56}} \Bigr)
 + {10\over 9} 
} \,,
\anoneqn
$$
$$
\hskip -.4 cm 
\eqalign{
F^\nn&(1,2,3,4)  = 
\biggl(
  {\spab3.{(1+2)}.6^2 \over \spa2.3 \spb5.6 t_{123} \spab1.{(2+3)}.4}
- {\spb1.2^2 \spa4.5^2 \over \spb2.3\spa5.6 t_{123} \spab4.{(2+3)}.1}
\biggr)\cr 
& \hskip 3 cm \times
\Bigl[ \Ls_{-1}\Bigl({-s_{12}\over -t_{123}}, {-s_{23} \over -t_{123}} \Bigr) 
     + \Lsnew^{2{\rm m}h}_{-1} (s_{34},t_{123};s_{12},s_{56} ) \Bigr] \cr
%
& \hskip 0.3 cm 
- 2 \, {\spab3.{(1+2)}.6 \over \spb5.6 \spab1.{(2+3)}.4 } \biggl[
    {\spab1.{(2+3)}.6 \spb1.2 \over t_{123} }
     {\Ll_0 \Bigl({-s_{23}\over -t_{123}}\Bigr) \over t_{123}}
 + {\spab3.4.6 \over \spa2.3}
     {\Ll_0 \Bigl({-s_{56} \over -t_{123}} \Bigr) \over t_{123} } 
     \biggr] \cr
& \hskip 0.3 cm 
- {1\over 2} { 1 \over \spa2.3 \spb5.6 t_{123}
       \spab1.{(2+3)}.4 } \biggl[
  \bigl( \spab3.2.1 \spab1.{(2+3)}.6 \bigr)^2
         { \Ll_1 \L {-t_{123}\over -s_{23}} \R \over s_{23}^2}
+ \spab3.4.6^2 t_{123}^2 
         { \Ll_1 \L{-s_{56} \over -t_{123}} \R \over t_{123}^2}
         \biggr] \cr\cr 
& \hskip 2 cm 
 - \hbox{flip} 
\,, \cr 
}\anoneqn
$$
where `flip' is to be applied to all the preceding terms in $F^\nn$.


\subsection{The Helicity Configuration $q^+\Qb^-Q^+\qb^-$}
\vskip5pt

We now give the result for $A_6^\an(1,2,3,4)$, which contributes to the 
leading color part of the partial amplitude
$A_{6;1}(1_\q^+,2_\Qb^-,3_\Q^+,4_\qb^-;5_\eb^-,6_e^+)$.
This amplitude is odd under the same flip symmetry~(\use\FlipSymmetry)
as $A_6^\nn$.  The tree amplitude is 
$$
A^{{\rm tree}, \, \an}_6(1,2,3,4) =
 -i \, {\spb1.3 \spa5.4 \spab2.{(1+3)}.6 \over s_{23} s_{56} t_{123}}
 -i \, {\spa2.4 \spb6.1 \spab5.{(2+4)}.3 \over s_{23} s_{56} t_{234}} \,.
\anoneqn
$$
Note that $A^{{\rm tree}, \, \an}_6(1,2,3,4) = -A^{{\rm tree}, \, \nn}_6(1,3,2,4)$.

For the one-loop contributions we have 
$$
V^\an(1,2,3,4) =  
- {1\over \eps^2} \L \L {\mu^2 \over -s_{12}}\R^\eps+ 
                            \L {\mu^2 \over -s_{34}}\R^\eps \R
 + {2\over 3 \eps} \L{\mu^2 \over -s_{23}}\R^\eps
 - {3\over 2} \ln\Bigl({-s_{23}\over -s_{56}} \Bigr)   
 + {10\over 9} \,.
\anoneqn 
$$
Note that $V^\an(1,2,3,4) = V^\nn(1,2,3,4)$.  The finite part is
$$
\hskip -.3 cm 
\eqalign{
F^\an&(1,2,3,4) = 
\biggl( -{\spb1.3^2 \spa4.5^2 \over
     \spb2.3 \spa5.6 t_{123} \spab4.{(2+3)}.1} 
+ {\spa1.2^2 \spab3.{(1+2)}.6^2
    \over \spa2.3 \spb5.6 t_{123} \spa1.3^2 \spab1.{(2+3)}.4} \Bigr) 
    \Ls_{-1}\Bigl({-s_{12}\over -t_{123}}, {-s_{23} \over -t_{123} } 
\biggr) \cr
& \hskip .3 cm
+ \biggl( -{\spb1.3^2 \spa4.5^2 \over
        \spb2.3 \spa5.6 t_{123} \spab4.{(2+3)}.1 } 
+ {\spab3.{(1+2)}.6^2 \spab2.{(1+3)}.4^2 \over
   \spa2.3\spb5.6 t_{123} \spab1.{(2+3)}.4 \spab3.{(1+2)}.4^2} 
  \biggr)
 \Lsnew^{2 {\rm m} h}_{-1} (s_{34},t_{123};s_{12},s_{56}) \cr
%
& \hskip .3 cm 
+ \biggl[ 
   {1\over 2} { (s_{12}-s_{34}) \spa5.1 \spab5.{(3+4)}.2 \spab1.{(2+4)}.3
     \over \spa5.6 \spab1.{(2+3)}.4 \spab1.{(3+4)}.2^2 } 
- {1\over 2} {(s_{12}-s_{34}) \spb2.3 \spa5.1 \spa5.2
   \over \spa5.6 \spab1.{(2+3)}.4 \spab1.{(3+4)}.2 } \cr
& \hskip 1 cm 
- {1\over 2} {(t_{134}-t_{234}) \spb3.4 \spa5.1\spa5.4
   \over \spa5.6 \spab1.{(2+3)}.4 \spab1.{(3+4)}.2}
- {\spb1.2 \spa5.1\spa5.2\spab1.{(2+4)}.3 
   \over \spa5.6 \spab1.{(2+3)}.4 \spab1.{(3+4)}.2  } \cr
& \hskip 1 cm 
+ {1\over 2} {\spa5.1 \spb6.2 \spab1.{(2+4)}.3 (t_{234}-2s_{34} )
   \over \spab1.{(2+3)}.4 \spab1.{(3+4)}.2^2 }
+ {1\over 2} {\spab5.{(1+2)}.6  \spab1.{(2+4)}.3 
    \over \spab1.{(2+3)}.4 \spab1.{(3+4)}.2} \cr 
&\hskip 1 cm     
- { \spab1.{(2+4)}.3 \bigl(
    s_{12} \d12(\spab5.{1}.6 - \spab5.{2}.6)
     - \d56 (\spab5.1.2 \spab2.5.6
           - \spab5.6.1 \spab1.2.6 )  \bigr) 
  \over \spab1.{(2+3)}.4 \spab1.{(3+4)}.2 \Delta_3 }  \cr
& \hskip 1 cm 
+ {1\over 2} {\spab2.{(1+4)}.3 \spab5.{(1+2)}.6 \d56 \over
   \spab1.{(2+3)}.4 \Delta_3} \biggr] 
   I_3^{3{\rm m}}(s_{12},s_{34},s_{56})  \cr
%
%
& \hskip .3 cm 
+ {\spb1.3\spa1.2 \spab3.{(1+2)}.6^2
   \over \spb5.6 t_{123} \spa1.3 \spab3.{(1+2)}.4}
       {\Ll_0\Bigl( {-t_{123}\over -s_{12}} \Bigr) \over s_{12}}
- {1\over 2} {\spb1.3^2 \spa2.3 \spab1.{(2+3)}.6^2
      \over \spb5.6 t_{123} \spab1.{(2+3)}.4 }
        {\Ll_1\Bigl( {-t_{123}\over -s_{23} } \Bigr) \over s_{23}^2 } \cr
& \hskip .3 cm 
+ {\spb1.3 \spab1.{(2+3)}.6 \spab2.{(1+3)}.6
    \over \spb5.6 t_{123} \spab1.{(2+3)}.4}
        {\Ll_0\Bigl( {-t_{123}\over -s_{23}} \Bigr) \over s_{23}}
+ {\spb1.3 \spa1.2 \spab1.{(2+3)}.6 \spab3.{(1+2)}.6 
   \over \spb5.6 t_{123} \spa1.3 \spab1.{(2+3)}.4 }
        {\Ll_0 \Bigl( {-t_{123} \over -s_{23} } \Bigr) \over s_{23}} \cr
& \hskip .3 cm 
- {1\over 2} {\spb6.4^2 \spa4.2^2    \, t_{123} \over
    \spa2.3 \spb5.6 \spab1.{(2+3)}.4}
       {\Ll_1\Bigl( {-s_{56}\over -t_{123}} \Bigr) \over t_{123}^2}
  - {\spb6.4^2 \spa4.2 t_{123}
      \over \spb5.6 \spab1.{(2+3)}.4 \spab3.{(1+2)}.4}
        {\Ll_0\Bigl( {-t_{123} \over -s_{56}} \Bigr) \over s_{56} } \cr
& \hskip .3 cm 
- 2 \, { \spb6.4 \spa4.2 \spab2.{(1+3)}.6 \over 
     \spa2.3 \spb5.6 \spab1.{(2+3)}.4}
        {\Ll_0 \Bigl( {-t_{123}\over - s_{56}} \Bigr) \over s_{56} } \cr
& \hskip .3 cm 
+ {1\over \spab3.{(1+2)}.4 \spab1.{(3+4)}.2 \Delta_3} \Bigl[
  (s_{12} - s_{34})\Bigl( {\spa1.2 \spb6.1 \spb6.2 t_{123} \over \spb5.6}
                          +{\spb1.2 \spa5.1 \spa5.2 t_{124} \over \spa5.6}
             \Bigr) \cr
& \hskip 3 cm 
 + (2 s_{12} - \d56 )
       \spa5.2 \spab1.{(3+4)}.2\spb1.6 \cr
& \hskip 3 cm 
 + \bigl((s_{13}+s_{23}) (s_{23}+s_{24}) - 
                 s_{12} (s_{12} + s_{23} - s_{14}) \bigr)
     \spab5.{1}.6 \cr
& \hskip 3 cm 
 + \bigl((s_{14}+s_{24}) (s_{13} + s_{14}) 
               - s_{12}(s_{12}-s_{23}+s_{14}) \bigr)
     \spab5.{2}.6 \Bigr]
         \ln\Bigl({-s_{12}\over -s_{56}}\Bigr) \cr
& \hskip 2 cm 
 - \hbox{flip} 
   \,, \cr 
}
\anoneqn
$$
where `flip' is to be applied to all the preceding terms in $F^\an$.

\subsection{Subleading Color Primitive Amplitude}

Here we give the primitive amplitude $A_6^\sl(1,2,3,4)$, 
which contributes only at subleading order in $N_c$.  
The ``tree amplitude'' appearing in eq.~(\use\VFdecomp) is 
$$
A_6^{\tree,\sl}(1,2,3,4)
=
i\, {\spb1.3 \spa5.4 \spab2.{(1+3)}.6 \over s_{12} s_{56} t_{123}} - 
i\, {\spa2.4 \spb6.3 \spab5.{(2+4)}.1 \over s_{12} s_{56} t_{124}} \,.
\anoneqn
$$
Note that $A^{\tree, \sl}_6(2,3,1,4) = -A^{{\rm tree}, \, \nn}_6(1,2,3,4)$.

For the subleading-color primitive amplitude,
 it is convenient to introduce an `exchange' operation
where the $5,6$ fermion pair is exchanged with 
the $1,2$ fermion pair,
$$
\hbox{exchange:} \hskip 1 cm 
1 \leftrightarrow 6 \,, \hskip 1 cm  
2 \leftrightarrow 5 \,.
\eqn\ExchangeSymmetry
$$
The box-parent part of $A_6^\sl$ is even under this exchange.
It is also convenient to define a symmetry operation
`flip\_sl' (distinct from
the leading-color `flip'),
$$
\hbox{flip\_sl:} \hskip 8 mm
1\leftrightarrow 5\,, \hskip 0.4 cm  
2\leftrightarrow 6\,, \hskip 0.4 cm  
3\leftrightarrow 4\,, \hskip 0.4 cm 
\spa{a}.{b} \leftrightarrow \spb{a}.{b} \,, \hskip 0.4 cm 
\sandmm{a}.{b}.{c} \leftrightarrow \sandpp{a}.{b}.{c}=\sandmm{c}.{b}.{a} \,.
\eqn\FlipslSymmetry
$$
The singular contribution is 
$$
\eqalign{
V^\sl(1,2,3,4) & = 
\biggl[- {1\over \eps^2} \L{\mu^2 \over -s_{34}}\R^\eps 
  - {3\over 2\eps} \L {\mu^2 \over -s_{34}}\R^\eps - 4  \biggr]
\ +\ \biggl[ - {1\over \eps^2} \L{\mu^2 \over -s_{12}}\R^\eps 
  - {3\over 2 \eps} \L{\mu^2 \over -s_{12}}\R^\eps 
  - {7\over 2}\biggr] 
\,, \cr}
\anoneqn
$$
where the first bracket is from the `box parent' graphs in
fig.~\use\SubleadingSumFigure{a} and the second bracket is from the
`triangle parent' graphs in fig.~\use\SubleadingSumFigure{b}. 
In fact, the entire contribution
from the triangle parent diagrams is contained in the second bracket.
The finite contribution is
$$
\hskip -1 cm 
\eqalign{
F^\sl&(1,2,3,4)  = 
\biggl[ {\spb1.3^2 \spa4.5^2 \over \spb1.2 \spa5.6 t_{123} \spab4.{(1+2)}.3 } 
 - {\spab3.{(1+2)}.6^2 \spab2.{(1+3)}.4^2
    \over \spa1.2 \spb5.6 t_{123}\spab3.{(1+2)}.4^3} \biggr]
     \Lsnew^{2 {\rm m} h}_{-1}(s_{34},t_{123};s_{12},s_{56}) \cr
%
&\hskip .3 cm
+ T \, I_3^{3 \rm m}(s_{12}, s_{34}, s_{56}) 
%
+\vphantom{dummy}  \biggl[
  {1\over 2} {\spb6.4^2 \spa4.2^2  t_{123}
    \over \spa1.2 \spb5.6 \spab3.{(1+2)}.4}
            {\Ll_1\Bigl( {-s_{56} \over -t_{123}}\Bigr) \over  t_{123}^2 }
+ 2 \, {\spb6.4 \spa4.2 \spab2.{(1+3)}.6
   \over \spa1.2 \spb5.6 \spab3.{(1+2)}.4 }
             {\Ll_0\Bigl( {-t_{123} \over -s_{56}} \Bigr) \over s_{56} } \cr
& \hskip 1.3 cm 
- {\spa2.3 \spa2.4 \spb6.4^2  t_{123}
   \over \spa1.2 \spb5.6 \spab3.{(1+2)}.4^2}
             {\Ll_0\Bigl({- t_{123} \over -s_{56}} \Bigr) \over  s_{56} }
- {1\over 2} { \spa2.3 \spb6.4  \spab2.{(1+3)}.6
    \over \spa1.2 \spb5.6 \spab3.{(1+2)}.4^2 }
             \ln\Bigl({- t_{123} \over -s_{56} } \Bigr)  \cr
& \hskip 1.3 cm 
- {3\over 4} {\spab2.{(1+3)}.6^2  
    \over \spa1.2 \spb5.6 t_{123} \spab3.{(1+2)}.4 }
             \ln\Bigl( {-t_{123} \over -s_{56} } \Bigr)
   - \hbox{flip\_sl} \biggr]  \cr
%
%
%
& \hskip .3 cm 
+ \vphantom{dummy} \biggl[  \ln\Bigl({-s_{12} \over -s_{34} }\Bigl)  \biggl(
  {3\over 2} {\spb1.2 \over \spab3.{(1+2)}.4 \Delta_3^2 }
     \bigl( \spb5.6 \spa2.5^2
         \d34\d56  
     - 2   \spa1.2^2 \spa5.6 \spb6.1^2   \d12
     + 4   \spa1.2 s_{56} \spa5.2 \spb6.1  \d56  \bigr) \cr
& \hskip 1.3 cm 
- {1\over 2} {\spb1.2 \spa2.5
              \over \spab3.{(1+2)}.4 \spa5.6 \Delta_3} 
     \bigl( \spa5.2 (s_{12}-s_{34}) - 2  \spa5.6 \spb6.1 \spa1.2 \bigr)  \cr
& \hskip 1.3 cm 
+ {3\over 2}  { t_{123} \over \spab3.{(1+2)}.4 \Delta_3^2}
     \bigl(\spa5.2 \spb1.6
            ( \d34\d56
             + 2  s_{12} \d12  ) 
     + 2 \, (\spb1.2 \spb5.6 \spa2.5^2 + \spa1.2 \spa5.6 \spb6.1^2)\,
           \d56  \bigr)  \cr
& \hskip 1.3 cm 
- {1\over 2} {t_{123} \over \spab3.{(1+2)}.4 \Delta_3} 
     \Bigl( \spa5.2 \spb1.6 
     - {\spb1.2 \spa2.5^2 \over \spa5.6} - 
       {\spa1.2 \spb6.1^2 \over \spb5.6} \Bigr) \cr
& \hskip 1.3 cm
+ \Bigl( - {\spb1.2 \spb6.4 \spa3.2 \over \spab3.{(1+2)}.4^2 \Delta_3}  
    + {\spab5.{(2+3)}.1 \over \spa5.6 \spab3.{(1+2)}.4 \Delta_3}  \Bigr)      
     (\spa5.2 \d56 - 2 \, \spab5.{(3+4)}.1 \spa1.2 )\cr
& \hskip 1.3 cm
- {\spb6.4 \spa3.2 t_{123} \over \spb5.6 \spab3.{(1+2)}.4^2 \Delta_3}         
     \bigl(\spb1.6 \d56 - 2 \, \spab2.{(3+4)}.6 \spb1.2 \bigr) \cr 
& \hskip 1.3 cm
+ {4\over \spab3.{(1+2)}.4 \Delta_3} 
       \bigl( \spa5.4 \spb4.1   \spab2.{(1+3)}.6
       + \spb6.3 \spa2.3   \spab5.{(2+4)}.1 \bigr)  \cr
& \hskip 1.3 cm
+ 2 \, {\spb6.4 \spa4.2 
   \over \spa1.2 \spb5.6 \spab3.{(1+2)}.4 \Delta_3  }
      \bigl( -2 \, \spab2.{5}.6 \d56
        + \spab2.{(3+4)}.6 \d34 \bigr) \cr
& \hskip 1.3 cm
- 2 \, {\spa5.2 \over \spa1.2 \spa5.6 \spab3.{(1+2)}.4 \Delta_3}
      \d34 \,
          \bigl( \spa5.3 \spb3.4 \spa4.2
          + \spa5.4 \spb4.1 \spa1.2
          - \spab5.{(1+2)}.3 \spa3.2 \bigr) \cr
& \hskip 1.3 cm
- {\spb6.4   \spab2.{(1+3)}.6   \spaa2.{(5+6)(1+2)}.3
   \over \spa1.2 \spb5.6 t_{123} \spab3.{(1+2)}.4^2 }
+ {1\over 2}  {\spa2.3 \spb4.6 \spab2.{(1+3)}.6
   \over \spa1.2 \spb5.6 \spab3.{(1+2)}.4^2 } \cr
& \hskip 1.3 cm
- {1\over 4}  {\spab2.{(1+3)}.6^2
    \over \spa1.2 \spb5.6 t_{123} \spab3.{(1+2)}.4 }
+  {\spa5.2   ( \spa4.5 \spab2.{(1+3)}.4 - \spa2.5 t_{123} )
  \over \spa1.2 \spa5.6 t_{123} \spab3.{(1+2)}.4   }     \biggr) 
   - \hbox{flip\_sl} \biggr]  
\cr 
%
& \hskip .3 cm 
- {1\over 2} {  ( t_{123} \d34 + 2 s_{12} s_{56})
  \over \spab3.{(1+2)}.4 \Delta_3}
    \Bigl( {\spb6.1^2 \over \spb1.2 \spb5.6} 
         + {\spa5.2^2 \over \spa1.2 \spa5.6} \Bigr)
+ (t_{123} - t_{124}) {\spb6.1 \spa5.2    
          \over \spab3.{(1+2)}.4 \Delta_3} \cr\cr
& \hskip 2 cm 
+ \hbox{exchange}
,\cr
} \anoneqn
$$
where `exchange' is to be applied to all the preceding terms 
in $F^\sl$, but `flip\_sl' is to be applied only to the terms 
inside the brackets ($[\ ]$) in which it appears. 
The three-mass triangle coefficient $T$ is given by
$$
\hskip -1 cm 
\eqalign{
 T & = 
 3 s_{34} (t_{123} \d34 + 2 s_{12} s_{56})
    { (\spb1.2\spa2.5^2\spb5.6 + \spa5.6\spb6.1^2 \spa1.2)
     \over \spab3.{(1+2)}.4 \Delta_3^2}  
- 6 s_{12} s_{56} s_{34} \spa5.2 \spb6.1
    { (t_{123}-t_{124})
         \over \spab3.{(1+2)}.4 \Delta_3^2}\cr
& \hskip .3 cm 
- {t_{123} \over \spab3.{(1+2)}.4 \Delta_3}
     ( \spa5.2 \spb6.1 (s_{56}+s_{12}+s_{34}) 
     + \spb1.2 \spa2.5^2 \spb5.6 + \spa5.6 \spb6.1^2 \spa1.2 ) \cr
& \hskip .3 cm             
-  {\spab5.{(2+3)}.1\over \spab3.{(1+2)}.4 \Delta_3}
      ( \spb6.5 \spa5.2 \d56 - \spb6.1\spa1.2 \d12 ) 
- {\spb1.2\spa5.6 \spab2.{(3+4)}.6^2 \over \spab3.{(1+2)}.4 \Delta_3}\cr
& \hskip .3 cm             
- {\spa2.3 \spb6.4 \over \spab3.{(1+2)}.4^2 \Delta_3}
    \bigl(\spa5.6 \d56 \, 
         (\spb6.1 t_{123} - \spb6.5 \spa5.2\spb2.1) 
     - \spb1.2 \d12
         (\spa2.5 t_{123} - \spa2.1 \spb1.6 \spa6.5 ) \bigr)   \cr
& \hskip .3 cm 
  - {\spa2.3 \spb6.4 \spab5.{(2+3)}.1 \over \spab3.{(1+2)}.4^2} 
- {1\over 2} 
 (t_{123} \d34 + 2s_{12} s_{56})
 {\spa2.3^2\spb6.4^2  \over \spa1.2 \spb5.6 \spab3.{(1+2)}.4^3 } \cr
%
& \hskip .3 cm 
- 2 \bigl( \spb6.5\spa5.2 \d56
      - \spb6.1 \spa1.2 \d12  \bigr) 
   {\spb6.3 \spa2.4 \over \spa1.2 \spb5.6 \Delta_3} \cr
& \hskip .3 cm 
+ 2 \, 
     \bigl( -\spab2.3.4 \spab4.5.6 \d56 
      + \spab2.1.3 \spab3.4.6 \d12
      + \spab2.{(1+3)}.6 s_{34} \d34 \bigr)
  {\spab2.{(1+3)}.6 
   \over  \spa1.2 \spb5.6 \spab3.{(1+2)}.4 \Delta_3} \cr 
& \hskip .3 cm 
+ 4 \, {\spab2.{(1+3)}.6 \spab5.{(2+3)}.1 s_{34}      
       \over \spab3.{(1+2)}.4 \Delta_3}
%
-  (t_{123} \d34 + 2 s_{12} s_{56})
    {\spab2.{(1+3)}.6 \spb1.4 \spa3.5 
        \over s_{12} s_{56} \spab3.{(1+2)}.4^2} \cr
& \hskip .3 cm 
+ 2 \,
      \bigl( - \spb1.2 \spa2.3 \spb3.4 \spa4.5
        + \spa5.6 \spb6.4 \spa4.3 \spb3.1
        + \spab5.{(2+3)}.1 s_{34} \bigr)
   {\spab2.{(1+3)}.6 
    \over s_{12} s_{56} \spab3.{(1+2)}.4 } \cr
& \hskip .3 cm 
- {1\over 2}  
     (t_{123} \d34 + 2s_{12} s_{56})
  { \spb1.6 \spa2.5
    \over s_{12} s_{56} \spab3.{(1+2)}.4 }  \cr
& \hskip .3 cm 
- {1\over 2}
       { (t_{123}-t_{124})
        \over s_{12} s_{56} \spab3.{(1+2)}.4 } 
        \bigl( 2\, \spab2.{(1+3)}.6 (\spab5.{3}.1-\spab5.{4}.1)\cr
& \hskip 4 cm 
          + (\spab2.{3}.6-\spab2.{4}.6) \spab5.{(2+4)}.1 
          + \spb6.1 \spa2.5 \d34 \bigr) \cr
& \hskip .3 cm 
+ 
     \biggl( - \spab5.{3}.1 + \spab5.{4}.1
     + {1\over 2}  (t_{123}-t_{124})
             \Bigl( {\spb1.4 \spa3.5 \over \spab3.{(1+2)}.4 }
                  + {\spb1.3 \spa4.5 \over \spab4.{(1+2)}.3 }\Bigr) 
     \biggr)
    {\spb3.6 \spa2.4 \over s_{12} s_{56}} \cr
\
& \hskip .3 cm 
-   \bigl( 4 \, \spab2.{(1+3)}.6 - \spab2.{3}.6 + \spab2.{4}.6 \bigr) 
   {\spb1.3 \spa4.5 \over  s_{12} s_{56}} 
  \,. \cr
}\anoneqn
$$

\subsection{Axial Vector Contribution}
\tagsubsection\AxialSection

The primitive amplitude $A_6^\ax(1,2,3,4)$ is unique in that neither of the 
external quark pairs couples directly to the vector boson; 
instead they couple through the fermion-loop triangle diagram 
shown in fig.~\use\AxialTriFigure. 
The contribution to the amplitude (\use\LoopColorDecomposition) 
vanishes when the vector boson is a 
photon (by Furry's theorem, i.e. charge conjugation
invariance).
The $Z$ contribution is proportional to the axial vector coupling of
the $Z$ to the quarks in the loop.  
As the $u,d,s,c$ quark masses may be neglected, only the $t,b$ quark pair survives 
an isodoublet cancellation in the loop, due to its
large mass splitting.  
We use the results of ref.~[\use\HHKY] to obtain this contribution;
that paper presents the fully off-shell $Zgg$ vertex, and we need only
contract it with the three fermion currents. 
The infrared- and ultraviolet-finite result is 
$$
\eqalign{ 
A_6^\ax(1,2,3,4)\ &=\ 
-{ 2 i \over (4\pi)^2 } 
{ f(m_t;s_{12},s_{34},s_{56}) 
  - f(m_b;s_{12},s_{34},s_{56}) \over  s_{56} }  
\biggl( { \spb6.3\spa4.2\spa2.5 \over \spa1.2 }
      - { \spb6.1\spb1.3\spa4.5 \over \spb1.2 } \biggr) \cr
&\qquad + (1\lr3, \hskip0.2cm  2\lr4)\ , \cr
}\eqn\Aaxformula
$$
where the integral $f(m)$ is defined in the 
appendix.
We need only the large mass expansion (for $m=m_t$) and the
$m=0$ limit (for $m=m_b$) of this integral;
these are presented in the appendix.

\section{Summary and Conclusions}
\tagsection\SummarySection

In this paper we presented all one-loop four-quark amplitudes which
enter into the computation of the next-to-leading order QCD 
corrections to $e^+e^- \to (\gamma^*,Z) \to 4$ jets.  
The two-quark two-gluon amplitudes will be
presented elsewhere [\use\Zqqgg,\use\ZqqggFuture]. 
We obtained the amplitudes by demanding that their functional forms
satisfy unitarity and factorization.  We also used a color
decomposition and a helicity basis to express the amplitudes in a
compact form.  This way of obtaining amplitudes is significantly more
efficient than using Feynman diagrams, since previously
computed amplitudes are used to construct new ones.
These amplitudes can be incorporated into numerical jet programs,
which should lead to an improved knowledge of the QCD background 
to searches for new physics in various processes. 
Indeed, the leading-in-$N_c$ parts of
the contributions presented here, together with those for 
two quarks and two gluons [\use\Zqqgg,\use\ZqqggFuture] have
already been inserted into one such program for four-jet
production in $e^+e^-$ annihilation [\use\Adrian].


\vskip0.3in
\par\noindent
{\bf Acknowledgements}
\vskip0.1in

L.D. thanks Adrian Signer for useful discussions.  
L.D. and D.A.K. are grateful for the support
of NATO Collaborative Research Grant CRG--921322.  S.W. acknowledges
the support of the Studienstiftung des deutschen Volkes and of
the CEA, under CFR \#161635.


\appendix{Integral Functions}

We collect here the integral functions appearing in the text,
which contain all logarithms and dilogarithms present in the 
amplitudes.  Except for the contribution of the top quark to the
fermion triangle loop in $A_6^\ax$, all internal lines are taken to 
be massless.  The following functions already appear in the evaluation
of pentagon loop integrals where all external legs are massless, 
and of box integrals with one external mass, e.g. as occur in the 
one-loop five-gluon amplitudes~[\use\FiveGluon]:
$$
\eqalign{
  \Ll_0(r) &= {\ln(r)\over 1-r}\,,\hskip 10mm
  \Ll_1(r) = {\Ll_0(r)+1\over 1-r}\,, \cr
  \Ls_{-1}(r_1,r_2) &=
      \Li_2(1-r_1) + \Li_2(1-r_2) + \ln r_1\,\ln r_2 - {\pi^2\over6}\,,\cr
}\eqn\Lsdef
$$
where the dilogarithm is
$$
\Li_2(x) = - \int_0^x dy \, {\ln(1-y) \over y}\,.
\eqn\Lidef
$$
The function $\Ls_{-1}$ is simply related to the scalar box integral 
with one external mass, evaluated in six space-time dimensions where 
it is infrared- and ultraviolet-finite.  

The box function analogous to $\Ls_{-1}$, but for two adjacent external
masses, is
$$
\eqalign{
  \Ls^{2{\rm m}h}_{-1}(s,t;m_1^2,m_2^2) &=
    -\Li_2\left(1-{m_1^2\over t}\right)
    -\Li_2\left(1-{m_2^2\over t}\right)
    -{1\over2}\ln^2\left({-s\over-t}\right)
    +{1\over2}\ln\left({-s\over-m_1^2}\right)
              \ln\left({-s\over-m_2^2}\right) \cr
&\quad
    + \biggl[ {1\over2} (s-m_1^2-m_2^2) + {m_1^2m_2^2\over t} \biggr]
        I_3^{3{\rm m}}(s,m_1^2,m_2^2) \,,\cr
}\eqn\Lstwomassdef
$$
where $I_3^{3{\rm m}}$ is the three-mass scalar triangle integral.
Here we use only a version of this box function 
with $I_3^{3{\rm m}}$ removed,
$$
\Lsnew^{2{\rm m}h}_{-1}(s,t;m_1^2,m_2^2) =
    -\Li_2\left(1-{m_1^2\over t}\right)
    -\Li_2\left(1-{m_2^2\over t}\right)
    -{1\over2}\ln^2\left({-s\over-t}\right)
    +{1\over2}\ln\left({-s\over-m_1^2}\right)
              \ln\left({-s\over-m_2^2}\right)\,.
\eqn\Lsnewdef
$$

The analytic properties of these integrals are straightforward to obtain
from the prescription of adding a small positive imaginary part to each
invariant, $s_{ij} \to s_{ij}+i\pol$.  One expands the logarithmic
ratios, $\ln(r) \equiv \ln({-s\over-s'}) = \ln(-s)-\ln(-s')$, and then 
uses
$$
  \ln(-s-i\pol)\ =\ \ln|s| - i\pi\Theta(s)\,.
\anoneqn  
$$
where $\Theta(s)$ is the step function: $\Theta(s>0) = 1$ and 
$\Theta(s<0) = 0$. 
The imaginary part of the dilogarithm $\Li_2(1-r)$ is given in terms of
the logarithmic ratio,
$$
  \Im\Li_2(1-r)\ =\ -\ln(1-r)\ \Im\ln(r)\, . 
\anoneqn  
$$
For $r>0$ the real part of $\Li_2(1-r)$ is given directly by 
eq.~(\use\Lidef).  For $r<0$ one may use~[\use\Lewin]
$$
  \Re\Li_2(1-r)\ =\  {\pi^2\over6} - \ln|r| \ln|1-r| - \Re\Li_2(r)\,, 
\anoneqn
$$ 
with $\Re\Li_2(r)$ given by eq.~(\use\Lidef).

The analytic structure of
$I_3^{3{\rm m}}$ is more complicated~[\use\ThreeMassTriangle,%
\use\UDThreeMassTriangle,\use\IntegralsLong], 
and the numerical representation we use 
depends on the kinematics.  The integral is defined by
$$
I_3^{3{\rm m}}(s_{12},s_{34},s_{56})\ =\ 
\int_0^1 d^3a_i\, \delta(1-a_1-a_2-a_3)\, 
{1 \over -s_{12}a_1a_2-s_{34}a_2a_3-s_{56}a_3a_1}\ .
\eqn\threemassintdef
$$
This integral is symmetric under any permutation of its three arguments,
and acquires a minus sign when the signs of all three arguments are  
simultaneously reversed.  Therefore we only have to consider two cases,
\par\noindent
1. The Euclidean region $s_{12},s_{34},s_{56}<0$, which is related by 
the sign flip to the pure Minkowski region ($s_{12},s_{34},s_{56}>0$) 
relevant for $e^+e^-$ annihilation.  Here the imaginary part vanishes.
This region has two subcases, depending on the sign of the 
Gram determinant $\Delta_3(s_{12},s_{34},s_{56})$ defined in
eq.~(\use\deltaijdef):
\par
1a. $\Delta_3<0$,
\par
1b. $\Delta_3>0$.
\par\noindent
2. The mixed region $s_{12},s_{56}<0$, $s_{34}>0$, for which $\Delta_3$
is always positive.  

In region 1a one may use a symmetric representation found by Lu and 
Perez~[\use\ThreeMassTriangle], which is closely related to that 
given in ref.~[\use\IntegralsLong]:
$$
\eqalign{
I_3^{3{\rm m}} &= {2\over\rtmdelta} \left[  
  \Cl_2\left(\! 2\tan^{-1}
       \left( { \rtmdelta \over \delta_{12} } \right) \! \right)
+ \Cl_2\left(\! 2\tan^{-1}
       \left( { \rtmdelta \over \delta_{34} } \right) \! \right)
+ \Cl_2\left(\! 2\tan^{-1}
       \left( { \rtmdelta \over \delta_{56} } \right) \! \right)
                                       \right] , \cr
}\eqn\LuPerezForm       
$$
where the $\delta_{ij}$ are defined in eq.~(\use\deltaijdef) and
the Clausen function $\Cl_2(x)$ is defined by
$$
\Cl_2(x)\ \equiv\ \sum_{n=1}^\infty { \sin(nx) \over n^2 }
 \ =\ -\int_0^x dt\, 
   \ln\bigl( \vert 2\sin(t/2) \vert \bigr).
\eqn\Cldef
$$

In regions 1b and 2 a convenient representation is given by 
Ussyukina and Davydychev~[\use\UDThreeMassTriangle],
$$
\eqalign{
I_3^{3{\rm m}}\ &=\ 
   -{1\over \rtdelta} \Re \left[  
    2 \left( \Li_2(-\rho x) + \Li_2(-\rho y) \right)
    + \ln(\rho x) \ln(\rho y)
    + \ln\left({y\over x}\right) 
         \ln\left({1+\rho y \over 1+\rho x}\right) 
    + {\pi^2\over3} \right] \cr  
&\qquad   
 - { i\pi \Theta(s_{34}) \over \rtdelta } 
   \ln\left( { (\d12+\rtdelta)(\d56+\rtdelta) \over
               (\d12-\rtdelta)(\d56-\rtdelta) } \right)\ , \cr
}\eqn\UDForm       
$$
where
$$
 x = {s_{12}\over s_{56}}, \quad y = {s_{34}\over s_{56}},
 \quad \rho = {2s_{56} \over \delta_{56}+\rtdelta}\,. 
\eqn\xydefs
$$

Finally, in the top quark contribution to $A_6^\ax$ the
combination $f(m_t)-f(m_b)$ appears, where $f(m)$ is the integral
$$
f(m;s_{12},s_{34},s_{56})\ = 
 \int_0^1 d^3a_i\, \delta(1-a_1-a_2-a_3)\, 
{a_2a_3 \over m^2-s_{12}a_1a_2-s_{34}a_2a_3-s_{56}a_3a_1}\ .
\eqn\fmdef
$$
This integral is complicated for arbitrary mass $m$; however,
the large and small mass limits of it suffice for $m_t$ and $m_b$
respectively.
For $m=m_t$ we simply Taylor expand the integrand in $1/m$;
for $m=m_b$ we set $m_b$ to zero, and reduce $f(0)$ to a linear
combination of the massless scalar triangle integral $I_3^{3{\rm m}}$ 
given above, logarithms and rational functions.
We get
$$
\eqalign{
&f(m_t;s_{12},s_{34},s_{56}) =  
 {1 \over 24 m_t^2 } + {(2s_{34}+s_{12}+s_{56}) \over 360 m_t^4 } 
 + \cdots\,, \cr
&f(0;s_{12},s_{34},s_{56}) =  
   \left( {3 s_{34} \d34 \over \dt^2} - {1 \over \dt} \right)
     s_{12} s_{56} I_3^{3{\rm m}}(s_{12},s_{34},s_{56}) 
  + \left( {3 s_{56} \d56 \over \dt^2} - {1\over2\dt} \right)
     s_{12} \ln\left({-s_{12}\over-s_{34}}\right)\cr
&\hskip 3 cm 
  + \left( {3 s_{12} \d12 \over \dt^2} - {1\over2\dt} \right)
     s_{56} \ln\left({-s_{56}\over-s_{34}}\right) 
  - { \d34 \over 2\dt }\,. \cr  
}\eqn\fmexplicit
$$

\listrefs
\listfigs

\end